\documentclass[aps,pre,twocolumn,showpacs,superscriptaddress]{revtex4-1}

\usepackage[dvipdfmx]{graphicx}
\usepackage[dvipsnames]{xcolor}
\usepackage{amsmath,amssymb,bm,amsthm,amsfonts}
\usepackage{physics}
\usepackage{lipsum}
\usepackage{braket}
\usepackage{ulem}
\usepackage{here}

\newcommand{\ignore}[1]{}
\newcommand{\mbold}[1]{\mbox{\boldmath $ #1 $}}

\newcommand{\beq}{\begin{equation}}
\newcommand{\eeq}{\end{equation}}

\def\SDG{{\rm \"{o}dinger equation }}

\begin{document}
\title{How to cross an energy barrier at zero Kelvin without tunneling effect} 

\author{Seiji Miyashita} 

\affiliation{Department of Physics, The University of Tokyo, 113-0033, Hongo, Bunkyo-ku, Tokyo, 113-0033, Japan}
\affiliation{The Physical Society of Japan, 2-31-22, Yushima, Bunkyo-ku, Tokyo, 113-0034, Japan}
\author{Bernard Barbara}
\affiliation{Institut N\'{e}el CNRS/UGA UPR2940
25 Avenue des Martyrs BP 166, 38042 Grenoble Cedex 9, France}

\date{\today}

\begin{abstract}
This Letter deals with the broad class of magnetic systems having a single or collective spin $S$ with an energy barrier, such as Rare-Earth elements and their compounds, Single Molecule Magnets with uniaxial anisotropy and more generally any other anisotropic quantum system made of single or multiple objects with discrete energy levels. Till now, the reversal of the magnetization of such systems at zero Kelvin 
required to make use of quantum tunneling with a significant transverse field or transverse anisotropy term, 
at resonance. Here, we show that another very simple method exists. It simply consists in the application of a particular sequence of electromagnetic radiations in the ranges of optical or microwave frequencies, depending on the characteristics of the system (spin and anisotropy values for magnetic systems).
This produces oscillations of the Rabi type that pass above the barrier, thus extending these oscillations between the two energy wells with mixtures of all the $2S+1$ states. 
In addition to its basic character, this approach opens up new directions of research in quantum information with possible breakthroughs in the current use of multiple quantum bits.
\end{abstract}

\maketitle

The magnetic properties of systems with uniaxial anisotropy are dominated by the presence of an energy barrier separating the spin-up and spin-down states. 
The reversal of a spin up/down to a spin down/up requires either a sufficient temperature for thermal activation above the barrier 
($T>T_{\rm B}=\Delta E/\ln(t/\tau_0)$, where $\Delta E$ is the energy barrier,
$\tau_0$ the usual prefactor of the Arrhenius law, and $t$ the time)
or the application of a sufficient transverse field to induce resonant tunneling through the barrier.

Demonstrated for the first time at the mesoscopic scale, this effect led to a stepwise hysteresis in the Single Molecule Magnet Mn$_{12}$-ac~\cite{SMM}. Later it was confirmed in many other mesoscopic systems (Fe$_8$~\cite{SMM-Fe8}, the Lanthanides
double-deckers~\cite{Lanthanide1, Lanthanide2}, etc.).
These results paved the way to the new field of magnetism at the mesoscopic scale (see e.g.,~\cite{SMM2}). 

In this Letter, we demonstrate the existence of a third possibility to pass a barrier as a simple consequence of the application of a special protocol of ac-magnetic fields,
a method which is valid with or without an applied magnetic field.
This protocol is purely quantum mechanical and aims to realize the control of the whole $2S+1$ states of a uniaxial spin $S$ without thermal excitation nor quantum tunneling. 
In particular, it shows that the coincidence of levels, required in quantum tunneling, 
is not required here.

\begin{figure}[H]
%\vspace*{-17mm}
\begin{center}
\hspace{10mm}\includegraphics[width=5cm]{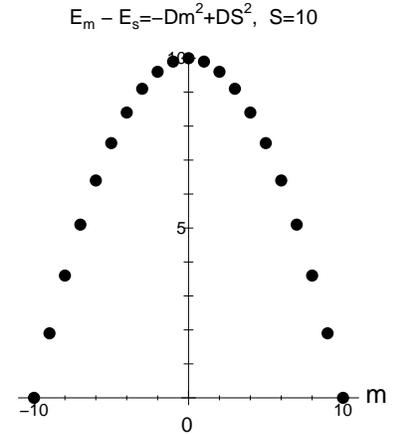}
\end{center}
%\vspace*{17mm}
\caption{Energies of states $E_m$ (Eq.(\ref{energy-m})) for $S=10$, $D=0.1$ and $H_z=0$.
We plot $E_m-E_{\rm min},\quad E_{\rm min}=E(\pm S)=-DS^2$,
showing the energy barrier between the states $m >0$ and $m <0$.
}
\label{energy-barrier}
\end{figure}

The dynamics of a spin $S$ of uniaxial anisotropy in a magnetic field Hz, is described by
\beq
{\cal H}=-DS_z^2-H_zS_z, \quad S_z=S, S-1,\cdots ,-S,
\label{Ham}
\eeq
where $D$ is the anisotropy constant. 
\noindent
The energy of the state $S_z=m$, given by
\beq
E_m=-Dm^2-H_zm,
\label{energy-m}
\eeq
enables to plot the energy barrier, 
\beq
E_m-E_{\rm min},\quad E_{\rm min}=-DS^2-H_zS,
\eeq
which is shown in Fig.~\ref{energy-barrier} for $S=10$ and $H_z=0$. 
When $D\ne 0$, the energy difference between the states
of $S_z=m-1$ and $S_z=m$ for a given field $H_z$ is given by
(for simplicity we take the Planck constant $\hbar=1$):
$$
\omega_{m\rightarrow m-1}\equiv E_{m-1}-E_m=H_z+D(2m-1),
$$ \beq
\quad m=S, S-1,\cdots -S+1.
\label{omega-m}
\eeq
If we add the circularly polarized frequency $\omega _{m\rightarrow m-1}$, the Hamiltonian~(\ref{Ham}) must be complemented by:
\beq
{\cal H}_{{\rm ac} m}=-h_{\rm ac}\left[\sin(\omega_{m\rightarrow m-1} t) S_x+
\cos(\omega_{m\rightarrow m-1} t) S_y\right],
\eeq
and the $z$-spin component oscillates between $m$ and $m-1$.
Note that, in the past, such state oscillations were used to prove the effect of photon-assisted quantum tunneling~\cite{Sorace}.

It is well known that in the limit $D=0$, the application of such a resonance ac-field in the $xy$ plane induces periodic spin oscillations between $S_z=S$ and $-S$,
a motion which is usually called Rabi oscillations (see Sec.~I in Supplementary Material (SM)~\cite{sec:SM}).
However, in the presence of an energy barrier $DS_z^2$, 
the Rabi oscillations are not possible because the level separations between the states $m$ and $m-1$ given by (\ref{omega-m}) are all different.
As we will see below, we will still create oscillations, applying several frequencies simultaneously between the states of
$S_z=m$ and $m-1$ for each separation (Sec.~II in SM~\cite{sec:SM}). 
\vspace*{-5mm}
\begin{figure}[H]
\vspace*{3mm}
\begin{center}
\includegraphics[width=6cm]{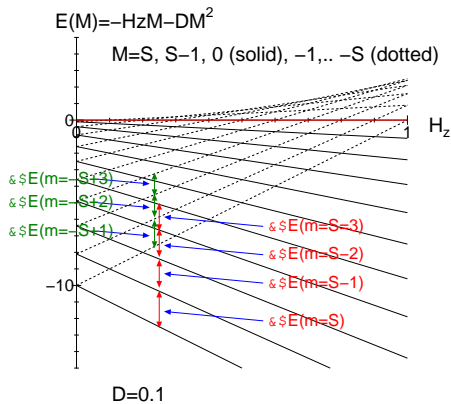}
\end{center}
%\vspace*{12mm}
\caption{Energy levels of the spin $S=10$ as a function of positive $H_z$ for $D=0.1$.
Some of energy difference between the states are depicted by arrows.}
\label{energy-levels}
\end{figure}
\vspace*{-1mm}
In the presence of a longitudinal field $H_z$, the energy levels, calculated for $S=10$ and $D=0.1$, are depicted Fig.~\ref{energy-levels}. 
The transition between each consecutive level ($E_m$ and $E_{m-1}$)
is induced by the application of a resonance field with 
the frequency $\omega_{m\rightarrow m-1}$ given by (\ref{omega-m}).

When we include the ac-fields up to $\omega_{S'\rightarrow S'-1}$, the Hamiltonian becomes:
$$
{\cal H}=-DS_z^2-H_zS_z
$$ 
\vspace*{-2mm}
\beq
-h_{\rm ac}\left(\sum_{m=S}^{S'}\sin(\omega_{m\rightarrow m-1} t)S_x
+\sum_{m=S}^{S'}\cos(\omega_{m\rightarrow m-1} t)S_y\right).
\label{Hamomega}
\eeq

\begin{figure}[h]
%\vspace*{-17mm}
\begin{center}
\hspace*{10mm}\includegraphics[width=5cm]{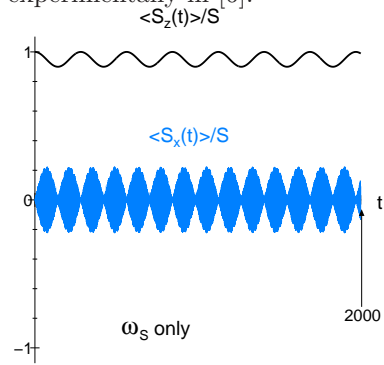}
\end{center}
%\vspace*{17mm}
\caption{Time dependence of the reduced spin components $S_x(t)/S$ (below, thin line) and 
$S_z(t)/S$ (above, bold line) under the ac-field 
of $\omega_{S\rightarrow S-1}$ only. $h_{\rm ac}=0.005$.
(In the figure $\omega_S$ denotes $\omega_{S\rightarrow S-1}$.)
The initial state is 
$(S_x/S,S_y/S,S_z/S)=(0,0,1)$.
As we have $S=10$, this scheme shows that $S_x$ oscillates between $+2$ and $-2$ and $S_z$ between $S{=10}$ and $S-1=9$.}
\label{Rabi-k=1}
\end{figure}

First, let us consider the case with a single $\omega_S$, which causes the oscillation between 
$S$ and $S-1$ (that is $S'=S$ in (\ref{Hamomega})) as shown in Fig.~\ref{Rabi-k=1}.
As expected, the result is a slow oscillation of $S_z$ between $m=S$ and $S-1$
with a rapid precession of the $x$-component around the $z$-axis.
In the particular case where the $S-1$ level is in coincidence with a level on the other side of the barrier, photon-activated tunneling should be observed, as this was confirmed experimentally in \cite{Sorace}.

\begin{figure*}[h]
%\vspace*{-17mm}
$$ \begin{array}{ccc}
\includegraphics[width=5cm]{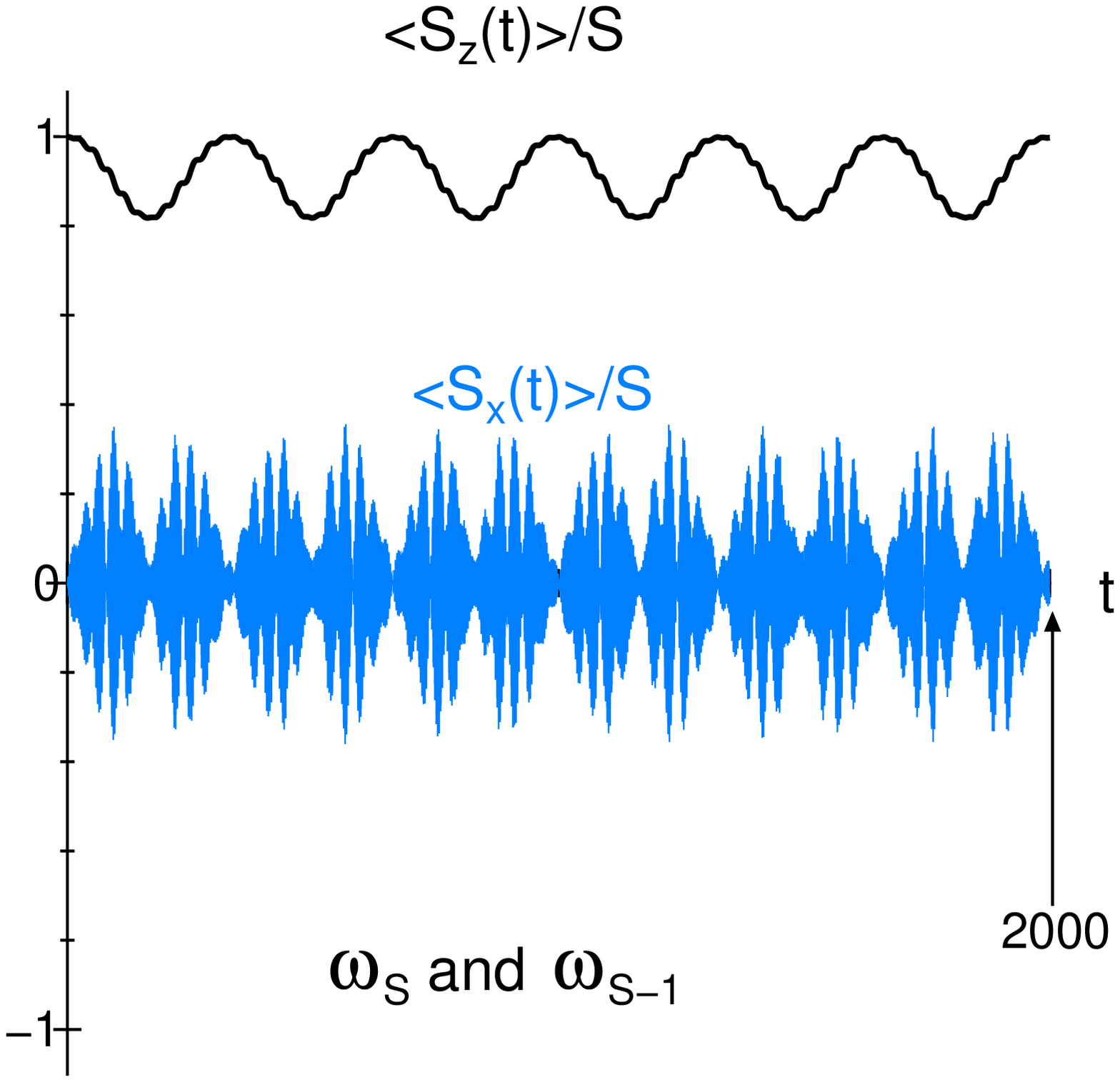}&
\includegraphics[width=5cm]{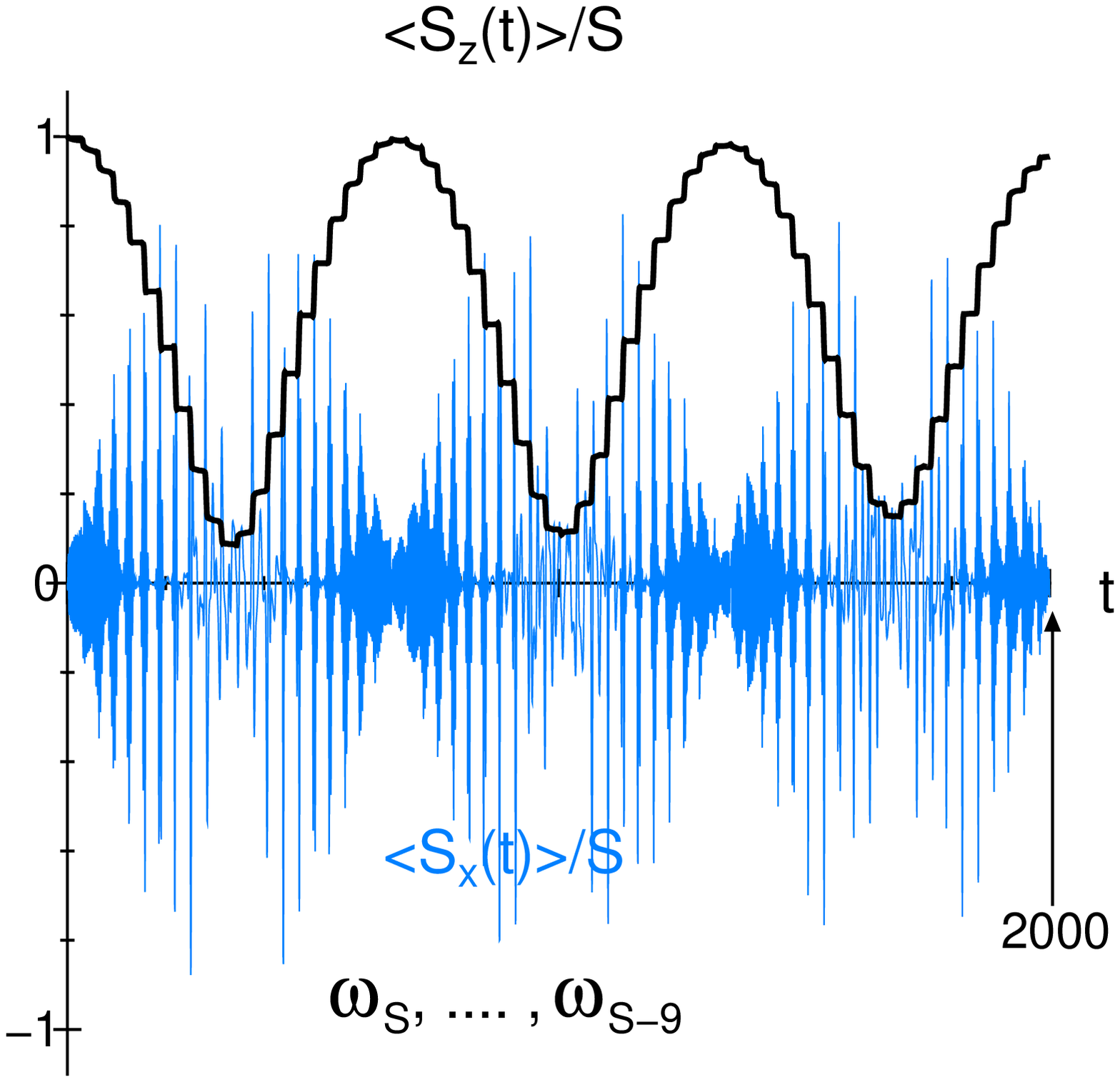}&
\includegraphics[width=5cm]{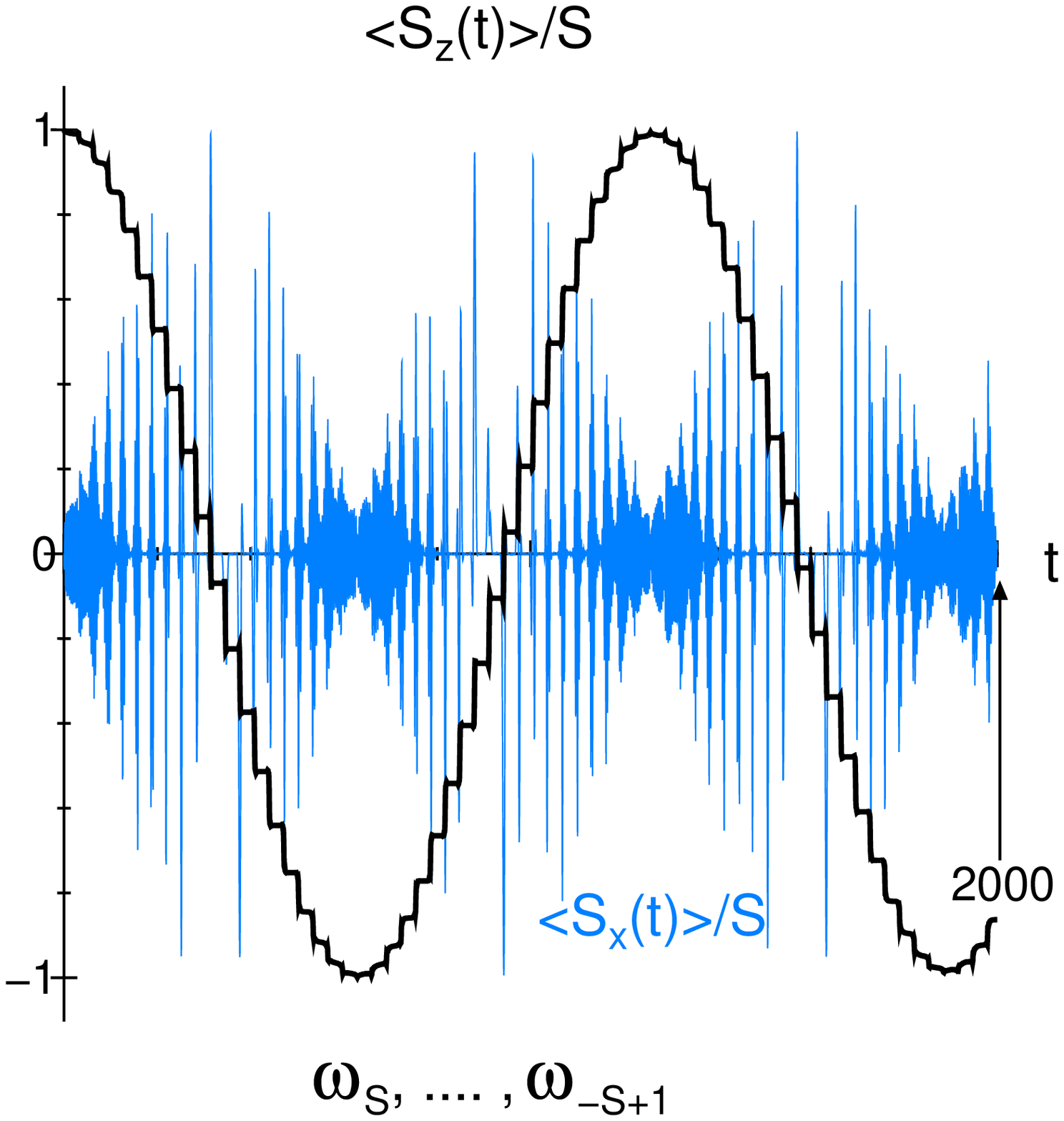}\\
{\rm (a)} & {\rm (b)} & {\rm (c)}\end{array}
$$
%\vspace*{17mm}
\caption{Time dependence of the reduced spins components $S_x(t)/S$ (blue thin line) and 
$S_z(t)/S$ (black bold line) under the ac-field with $h=0.005$ and $H_z=0$.
(a) $\omega_{S\rightarrow S-1}$ and $\omega_{S-1\rightarrow S-2}$ and
(b) $\omega_{S\rightarrow S-1},\cdots \omega_{S-9\rightarrow S-10}$.
(c) $\omega_{S\rightarrow S-1},\cdots \omega_{-S+1\rightarrow S}$.
In all the cases, the initial state is $(S_x,S_y,S_z)=(0,0,1)$.
}
\label{Rabi-k=2}
\end{figure*}

If we now include one more ac-field, i.e., $\omega_{S-1\rightarrow S-2}$ on the top of $\omega_{S\rightarrow S-1}$, we find
the superposition of two successive resonances: 
one between the levels $m=S$ and $S-1$ and one between $S-1$ and $S-2$ (Fig.~\ref{Rabi-k=2}(a)). 
This leads to 
more complex oscillations of $S_z$ between $m=S$ and $S-2$, together with a rapid precession of the $x$-component around the $z$ axis. 
Interestingly,
such a spin motion corresponds to nothing else but an incomplete Rabi oscillation,
in which the $xy$ components are modulated with an envelope corresponding to the period of the $z$ component.

Taking now all the ac-fields of one side of the barrier (from $\omega_{S\rightarrow S-1}$ to 
$\omega_{S-9\rightarrow S-10}$), 
we find similar oscillations of the $x$ and $z$ components with, however, larger amplitudes and modulations. In particular, $S_z$ oscillates between 
the ground-state $S_z=10$ and the top of the barrier (Fig.~\ref{Rabi-k=2}(b).
(All cases of $S'=S, S-1,\cdots -S+1$ are shown Fig.~3 of SM~\cite{sec:SM})

Finally, if we include all the resonance fields given by $\omega_{m\rightarrow m-1}$ (\ref{omega-m}), i.e., $S'=-S+1$, 
we find that the spin oscillations extend above the barrier between the states $S$ 
and $-S$ (Fig.~\ref{Rabi-k=2}(c)). 
The interferences between the corresponding oscillations associated with all the $x$, $y$ and $z$ spin components of each one of the 21 states of our spin $S=10$, lead to rather fast in-plane spin oscillations with much slower, and also almost complete perpendicular oscillations between $S$ and $-S$.
Even if it is purely quantum, this motion is rather similar to, 
but quite different from a classical Rabi oscillation. We call it ``Giant Quantum Oscillations Above the Barrier" (GQOAB).
Note that the present oscillations from one well to the other one differ from those resulting from quantum tunneling because the application of a significant transverse field and the coincidence of the spin-up and spin-down levels are not required. 

\onecolumngrid

\newpage
\twocolumngrid

In order to show some more details regarding the properties of our GQOAB,
we write the Schr\"{o}dinger equation in the rotating frame (Sec. III-A in SM~\cite{sec:SM}):
\beq
i\hbar{\partial \over\partial t}|\Phi(t)\rangle=
\left(-DS_z^2-h_{\rm ac}{f(t)\over 2i}\left(S^+-S^-\right)\right)|\Phi(t)\rangle,
\label{dPhidt}
\eeq
where the sinusoidal functions of (\ref{Hamomega}) can be written
\beq
f(t)={\sin(2DSt)\over\sin(Dt)},
\label{ft2DtSDt}
\eeq
which is depicted in SM (Fig.~4 of SM~\cite{sec:SM}).
These equations do not depend on $H_z$, contrary to the case of quantum tunneling where the spin-up and down states must be in coincidence (see Fig.~5 of SM~\cite{sec:SM}).

This new mechanism induces spin reversal above the barrier as this is the case with thermal activation, but here the activation is coherent and associated with the application of a particular time-dependent electromagnetic field $f(t)$ at zero Kelvin.

\begin{figure}[h]
%\vspace*{-17mm}
$$
\begin{array}{c}
\hspace*{10mm}\includegraphics[width=5cm]{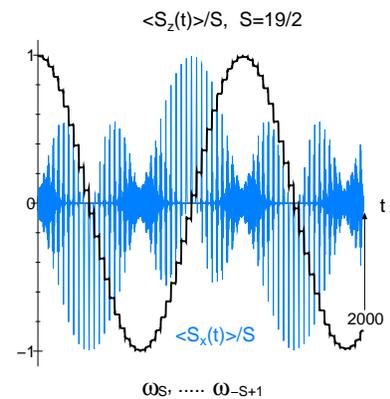}\\
\end{array}
$$
%\vspace*{17mm}
\caption{Time dependence of the reduced spin components $S_x(t)/S$ (below, thin line) and $S_z(t)/S$ (above, bold line) under the ac-field of $\omega_{S\rightarrow S-1},\cdots \omega_{-S+1\rightarrow -S}$
for a non-integer spin $S=19/2$. $H_z=0$ and $h_{\rm ac}=0.005$.
(In the figure $\omega_S$ represents $\omega_{S\rightarrow S-1}$, and so on.)
The initial state is $(S_x,S_y,S_z)=(0,0,1)$.}
\label{Rabi-k=19}
\end{figure}
Clearly enough, these calculations performed with the example of a spin $S=10$, which corresponds to the case of Mn$_{12}$-ac, could be done with any spin size. 
For example, as the calculations are done at zero Kelvin, we could also have taken the smallest possible spin with a barrier, $S = 1$. However, on the experimental
side, this would be more tricky because the height of the
barrier being $DS^2$, the use of very small spins would require extremely low temperatures, often not available. 
Regarding the high spins side, the limitation comes from the fact that the spin levels become very close to each other making difficult the applications of the adapted microwaves frequencies.
We have shown that this GQOAB is also valid for any other integer and also non-integer spin, even if the energy structure is slightly different for the non-integer case
(see Fig.~\ref{Rabi-k=19} for $S=19/2$ as an example in non-integer case, and also Sec.III-B in SM~\cite{sec:SM} for $S=5$).

\begin{figure}[h]
%\vspace*{-17mm}
$$ \begin{array}{c}
\hspace*{10mm}\includegraphics[width=5.5cm]{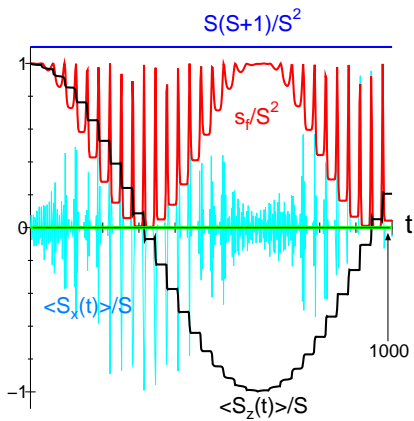}
\end{array}
$$
%\vspace*{17mm}
\caption{Motions of spin $S=10$ for $H_z=0.1$, $h_{\rm ac}=0.005$ and the driving ac-field $f(t)$
given by Eq.~(\ref{ft2DtSDt}). 
The amplitude $h_{\rm ac}=0.005$.
$\langle S_x(t)\rangle$ (thin curve) and $\langle S_x(t)\rangle$(bold curve) are normalized by $S$, 
The normalized spin length (spin-fidelity $s_f$) is given by the bold line (above).
The total spin $S(S+1)$ normalized by $S^2$ is given by the blue line.
}
\label{figCRabi20}
\end{figure}
Despite their similarities, our GQOAB and the classical Rabi oscillations are obviously very different, 
the first being quantum with finite $D$ while the second can be classical with $D=0$ (Fig.~2 in SM~\cite{sec:SM}).
For example, in our GQOAB, the spin-fidelity\cite{Hatomura2016,SM2022}: 
\beq
s_f=\langle S_x\rangle^2+\langle S_y\rangle^2+\langle S_z\rangle^2
\eeq
is not conserved, which is obviously in opposition with the case of classical Rabi oscillations (see Fig.~\ref{figCRabi20} and Sec.~I of SM~\cite{sec:SM}). 
In contrast, the total spin
$
\langle S_x^2\rangle+\langle S_y^2\rangle+\langle S_z^2\rangle=S(S+1)$ 
is conserved in both cases. 

It is also noted that when $D=0$, all the resonance frequencies become the same, i.e., $\hbar\omega_{m\rightarrow m-1}=H_z$. 
In this case, the amplitude of the ac-field is $2Sh_{\rm ac}$, and 
$\langle S_z\rangle$ shows usual Rabi oscillations with a period proportional to $1/2Sh_{\rm ac}$. 
Conversely, for large $D$ this period is simply proportional to $1/h_{\rm ac}$ showing that, when $D$ approaches 0, the period of GQOAB oscillations changes very rapidly
in a complicated manner especially when $S$ is large (see Fig.~9(c) in Sec.~III-D in SM~\cite{sec:SM}). 

Our very simple procedure to create quantum oscillations going through all the states from one side of the barrier to the other one, could be realized with single molecule magnets or 3d or 4f elements, 
highly diluted in a non-magnetic matrix single crystal.
Single-domain nanoparticles could also be used, if the required important set of frequencies is available. 
As an example we might suggest the case of Ho$^{3+}$ ions diluted in the matrix of YLiF$_4$~\cite{YLiF4} in which a barrier evidenced by a Mn$_{12}$-like stepwise hysteresis loop was observed. Furthermore, this system offers the possibility to extend our approach when the electronic levels on each side of the barrier are splitted by hyperfine interactions.

In conclusion, through the application of a particular protocol of experimentally accessible electro-magnetic radiations, we found a new method to reverse a spin above a barrier at zero Kelvin, whatever its parity is.
This mechanism, which does not require the application of a transverse field or the coincidence of spin-up and spin-down states, is not a tunneling effect, even if it shows similar quantum oscillations between the ground spin-up and spin-down states. In our case, the frequency of these oscillations is proportional to the amplitude of the applied electromagnetic radiation field.
We have demonstrated the method with $S=10$. In this case, we need 20 different frequencies while a spin $S=20$ would require 40. 
Using a EPR spectrometer generating the so-called ``shaped pulses"~\cite{NN}, we are currently performing experiments that will enable to elucidate 
what the limitations on the size of large spins are.

Together with its basic interest, which gets back to the early days of magnetism by showing how to pass a barrier at zero Kelvin without using the tunnel effect, GQOAB opens new possibilities in quantum information through very simple spin manipulations in the presence of a barrier. 
This procedure is ``active'' as it leads to controllable oscillations over the barrier between two arbitrary states, unlike the quantum tunneling effect where, since 
the levels must coincide, there is no way to monitor the quantum oscillations. 
In the present method, all the transitions between any pair of the $2S+1$ states are possible.
The energies of the initial state and the target state are not necessarily the same as 
it is the case with quantum tunneling. Furthermore, parasitic magnetic fields will not have any influence in e.g., decoherence as the transitions do not depend on the external magnetic field $H_z$.

Even if this was not our main motivation, this approach enabling the superposition of the whole $2S+1$ states of a spin on both sides of the energy barrier, should pave the way to new methods of quantum manipulations for quantum computing, transfer between different states, contiguous, noncontiguous, in a same barrier-well or in different barrier wells.

During the last decade, qubit manipulations of single-spins magnets with or without rare-earths, or with or without nuclear spins (see e.g., \cite{YLiF4}) have been extensively investigated together with other materials, such as single-electron quantum dots in superconducting circuits (see the review~\cite{Y} and therein references). 
More recently, special integrations mixing electro-nuclear spins qubits and quantum circuits have been realized~\cite{N6,N7,N8,N9,N10,N11} enabling the implementation of various error corrections quantum algorithms~\cite{N12,N13,N14,N15,N16}. 
In particular the multi-bits spins manipulations that are currently underway~\cite{N2} involve quantum states connection 
through the creation/annihilation of the ladder operators, as in our approach, but in the absence of a barrier which greatly reduces the possibilities of manipulation, in particular, because in the presence of an energy barrier, the entanglements are restricted to the pair of states which is in coincidence.
By contrast, the use of the precisely shaped microwave frequencies proposed in this paper, should permit the simultaneous quantum mechanical control of the 2S + 1 states in the presence of a barrier and should therefore provide a novel generic strategy for the integration of various quantum magnetic systems.
Furthermore, the populations of the states separated by a barrier are expected to be more stable against decoherence, which is particularly true here, as the Hamiltonian is independent of $H_z$. 

In order to obtain GQOAB at finite temperature, the temperature must be sufficiently low for the microwave-induced jump probability to be
much higher than the thermal activation jump probability.
Taking the example of Mn$_{12}$-ac with $S = 10$ and $D \simeq 0.6$K, we find $T\ll 15$K
which is more than acceptable. 
In addition, decoherence effects~\cite{CL,Spinbath} 
should be lower than with Rabi oscillations because GQOAB are (i) non-sensitive to external magnetic fields and (ii) continuously supported by applied microwaves. 
And, as the Rabi oscillations are now reaching coherence times of the order of the microseconds\cite{Z1,Z2} and even of the dozen microseconds~\cite{Irinel-Sylvain} (in the example of CaWO4:Gd) we expect that the use of GQOAB will not be limited by decoherence effects.

This work was supported by the Elements Strategy Initiative Center for Magnetic Materials (ESICMM) (Grant No. 12016013) funded by the Ministry of Education, Culture, Sports, Science and Technology (MEXT) of Japan, and was partially supported by Grants-in-Aid for Scientific Research C (No. 18K03444 and No. 20K03809) from MEXT. S.M. acknowledges the hospitality of ``JSR-UTokyo Collaboration Hub, CURIE''.

\newpage
\onecolumngrid

\begin{center}
{\Large Supplementary Material}
\end{center}

\section{Transition in a two-state system with a resonance field (Rabi oscillation)}\label{sec:Rabi}
It is well known that the transition between two states with the energy difference $\Delta E$ 
can be induced by a resonance field.
For clarity, we will take a brief look at the mechanism below. 
Let us consider the two states A and B and 
their energies $E_{\rm A}$ and $E_{\rm B}$, with $E_{\rm A} <E_{\rm B}$.
If we assign the states A and B of a spin to the eigenstates
$S_z= 1/2$ and $S_z= -1/2$, the Hamiltonian becomes:
\beq
{\cal H}=-H_z S_z +E_0,\quad \Delta E=H_z=E_{\rm B}-E_{\rm A},\quad E_0={E_{\rm B}+E_{\rm A}\over 2}.
\eeq
Hereafter we put $E_0=0$ for simplicity.

\begin{figure}[h]
$$ \begin{array}{c}
\includegraphics[width=7cm]{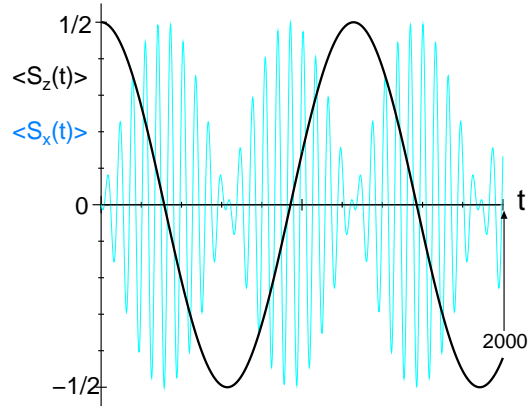}
\end{array}
$$
\caption{Oscillation of a spin ($S=1/2$) under the static field $H_z=0.1$, the ac field of the resonance frequency $\hbar\omega=H_z$ and the amplitude $h_{\rm ac}=0.005$.
$\langle S_x(t) \rangle$ (thin curve) and $\langle S_x(t) \rangle$(bold curve) normalized by $S$. 
}
\label{Rabi000}
\end{figure}

If we apply a circularly polarized field:
\beq
{\cal H}_{{\rm ac}}=-h_{\rm ac}(\sin(\omega t) S_x+\cos(\omega_t) S_y),
\label{resonancefield}
\eeq
with the amplitude $h_{\rm ac}$,
the Hamiltonian and corresponding Schr\"{o}dinger equation become:
\beq
{\cal H}=-H_z S_z -h_{\rm ac}\left(\sin(\omega t)S_x+\cos(\omega t)S_y\right),
\eeq
and 
\beq
i\hbar{d\over dt}\Psi(t)=\left(-H_zS_z -h_{\rm ac}\left(\sin(\omega t)S_x+\cos(\omega t)S_y\right)\right)\Psi(t).
\label{Seq}
\eeq
Now we consider the state in the rotation frame $\Phi(t)$ with the frequency $\omega$:
\beq
\Psi(t)=e^{i\omega S_zt}\Phi(t).
\label{Psi}
\eeq
Then, the left-hand side of Eq.(\ref{Seq}) is 
\beq
i\hbar{d\over dt}e^{i\omega S_zt}\Phi(t)=i\hbar\left(i\omega S_ze^{i\omega S_zt}\Phi(t)
+e^{i\omega S_zt}{d\over dt}\Phi(t)\right)
=-\hbar\omega S_ze^{i\omega S_zt}\Phi(t)+i\hbar e^{i\omega S_zt}{d\over dt}\Phi(t),
\eeq
and 
the Schr\"{o}dinger equation({\ref{Seq}}) is written as
\beq
i\hbar {d\over dt}\Phi(t)
=(\hbar\omega-H_z) S_z\Phi(t)-
h_{\rm ac}e^{-i\omega S_zt}\left(\sin(\omega t)S_x+\cos(\omega t)S_y\right)e^{i\omega S_zt}\Phi(t).
\label{dphidt}
\eeq
Using the following relations for the ac-field part:
\beq
e^{-i\omega S_zt}S^+e^{i\omega S_zt}=e^{-i\omega t}S^+,\quad 
e^{-i\omega S_zt}S^-e^{i\omega S_zt}=e^{i\omega t}S^-,
\label{SpSmomega}
\eeq
Rewritting the ac-field part as
\beq
\sin(\omega t)S_x+\cos(\omega t)S_y=
{e^{i\omega t}-e^{-i\omega t}\over 2i}{S^++S^-\over2}+{e^{i\omega t}+e^{-i\omega t}\over 2}{S^+-S^-\over 2i}={e^{i\omega t}S^+-e^{-i\omega t}S^-\over 2i},
\eeq
and using the relations (\ref{SpSmomega}), the second term of (\ref{dphidt}) becomes:
\beq
e^{-i\omega S_zt}\left(\sin(\omega t)S_x+\cos(\omega t)S_y\right)e^{i\omega S_zt}
=e^{-i\omega S_zt}\left({e^{i\omega t}S^+-e^{-i\omega t}S^-\over 2i}\right)e^{i\omega S_zt}
={S^+-S^-\over 2i},
\eeq
and then the equation (\ref{dphidt}) is given by
\beq
i\hbar {d\over dt}\Phi(t)=\left(-(H_z-\hbar\omega)S_z-h_{\rm ac}S_y\right)\Phi(t),
\eeq
where the Hamiltonian in the rotating frame is:
\beq
{\cal H}_{R}=-(H_z-\hbar\omega)S_z-h_{\rm ac}S_y.
\label{HRDeltaE}
\eeq
and the motion in the rotating frame is
\beq
\Phi(t)=e^{i((H_z-\hbar\omega)S_z+h_{\rm ac}S_y)t/\hbar}\Phi(0).
\eeq
This is a precession, in the rotating frame, around the field:
\beq
\mbold{H}=(0,h_{\rm ac},H_z-\hbar\omega).
\eeq
At resonance 
\beq
\hbar\omega=H_z, 
\eeq
the Hamiltonian and the corresponding wave function in the rotating frame become 
\beq
{\cal H}_{R}=-h_{\rm ac}S_y, \quad \Phi(t)=e^{ih_{\rm ac}S_yt/\hbar}\Phi(0),
\eeq
which is a rotation around the $y$-axis.
Here the $z$-component $\langle S_z\rangle$ changes sinusoidally in time with the period
\beq
T={2\pi\over h_{\rm ac}},
\eeq
while the $x$-component rotates much faster (Fig.~\ref{Rabi000}).

This motion, called Rabi oscillation, is often used to create the $\pi$-pulse of the spin-echo method, which corresponds to one half-period $T/2$ of the Rabi oscillation.

\begin{figure}[h]
$$ \begin{array}{c}
\includegraphics[width=7cm]{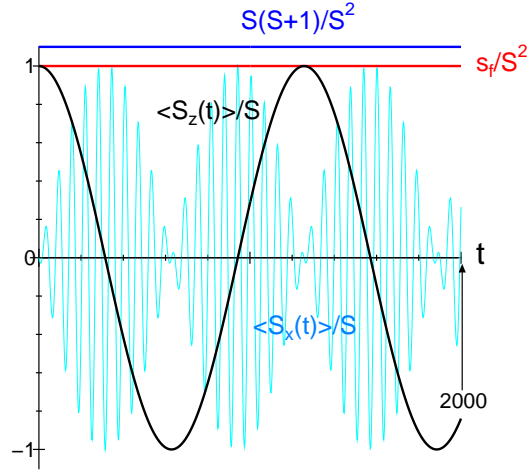}
\end{array}
$$
\caption{Motions of oscillation of a spin ($S=10$) under the static field $H_z=0.1$ and the ac field of the resonance frequency $\omega=H_z$ and the amplitude $h_{\rm ac}=0.005$.
$\langle S_x(t) \rangle$ (thin curve) and $\langle S_z(t) \rangle$(bold curve) normalized by $S$.
The normalized spin length (spin-fidelity (\ref{spin-fidelity})) is given of the red line.
The total spin $S(S+1)$ normalized by $S^2$ is given by the blue line.
}
\label{Rabi00}
\end{figure}

In Fig.~\ref{Rabi000}, we show the oscillation between $S_z=1/2$ and $S_z=-1/2$.
But, it should be noted that this oscillation does not depend on the value of $S$.
This is easily understood, because we do not use the value of $S$ in the above calculation. 
Figure~\ref{Rabi00} shows the dynamics of the system with $S=10$ in the static field $H_z=0.1$.

It is also noted that the motion of $\langle S_z(t)\rangle$ does not depend on $H_z$.
while the period of $\langle S_x(t)\rangle$ depends on $\omega=H_z/\hbar.$
This motion of the Rabi oscillations can be regarded as a rotation of magnetic vector of length $S$ around the $z$-axis, which is also the case of the classical spin.
In this case, the spin the spin-fidelity $s_f$ (T. Hatomura, B. Barbara and S. Miyashita,
Phys. Rev. Lett. {\bf 116} (2016) 037203). 
\beq
s_f=\langle S_x\rangle^2+\langle S_y\rangle^2+\langle S_z\rangle^2
\label{spin-fidelity}
\eeq
is constant in time, i.e.,
\beq
s_f=S^2,
\eeq
which is depicted by the red line in Fig.~\ref{Rabi00}.

\section{Transition between two states in a uniaxial spin system ${\bf (S>1/2)}$ with a resonance field (Resonance operation)}\label{sec:twostates}

Next, we study a transition between two states in a uniaxial spin $S$ of anisotropy $D$ consisting of more than two levels, e.g., a model with $S=10$ which has $(2S+1)$ states with $S_z=m=S, S-1,\cdots, -S$.
The corresponding Hamiltonian writes:
\beq
{\cal H}=-DS_z^2-H_zS_z, \quad S_z=S, S-1,\cdots ,-S.
\eeq
The energy of the state $S_z=m$ is
\beq
E_m=-Dm^2-H_zm,
\eeq 
while the energy barrier between the two lowest states $Sz=\pm S$ (see Fig.~1 in the main text)
writes:
\beq
E_{\rm barrier}=DS^2.
\eeq
In this system with anisotropy ($D\ne 0$), the single ac-field which we studied for $D=0$ in the previous section does not raise the spin reversal.

Now, we consider the transition between two sequential states of $S_z=m$ and $S_z=m-1$ by using the circularly polarized field with the resonance frequency. 
The energy difference between these states is
$$
\hbar\omega_{m\rightarrow m-1}\equiv E_{m-1}-E_m=-D(m-1)^2-H_z(m-1)-\left(-Dm^2-H_zm\right)
$$ \beq=H_z+D(2m-1),
\quad m=S, S-1,\cdots -S+1.
\label{omehamtom-1}
\eeq
Thus, an oscillation between the states is induced by the resonance ac-field (\ref{resonancefield})
\beq
{\cal H}_{{\rm ac}}=-h_{\rm ac}(\sin(\omega_{m\rightarrow m-1}t) S_x+\cos(\omega_{m\rightarrow m-1}t) S_y).
\label{resonancefield-omegam}
\eeq
When the initial state is at $S_z=S$, application of the ac-filed with 
$\omega_{S\rightarrow S-1}$ induces an oscillation between the states of $S_z=S$ and $S-1$
as shown in the text (Fig.~3 of the main text).

The operation $e^{-i{\cal H}_{{\rm ac}}t/\hbar}$
\beq
U(m,m-1,t)=e^{-i{\cal H}_{{\rm ac}}t/\hbar}
\eeq
causes the oscillation between the state of $m$ and $m-1$.
This operation also causes some effects on other states, but because of off-resonance, the effect is practically negligible on other pair of state. 
Thus, we may use this resonance field to make transitions between the states 
$m$ and $m-1$. (In case of $S=1/2$, the transition is exact, and for $S>1/2$, it is a good approximate transition.)
Indeed, this resonance oscillation is used in the magnetic resonance (ESR) to determine the energy level separation.

If we set the time to be the half-period 
\beq
t=T/2=\pi/h_{\rm ac}, \quad T=2\pi/h_{\rm ac},
\label{Period-pi}
\eeq 
the operation
\beq
X(m,m-1) =U(m,m-1,2\pi/h_{\rm ac})= e^{-i{\cal H}_{{\rm ac}}(\pi/h_{\rm ac})/\hbar} 
\label{Umm-1}
\eeq
plays the role of $\pi$-pulse and causes the transition between the states $m$ and $m-1$.
The operation $X(S,S-1)$ (\ref{Umm-1}) transfers the state ($|S\rangle$) of $S_z=S$ to the state ($|S-1\rangle$) of $S_z=S-1$.

\begin{figure}[h]
$$ \begin{array}{c}
\includegraphics[width=7cm]{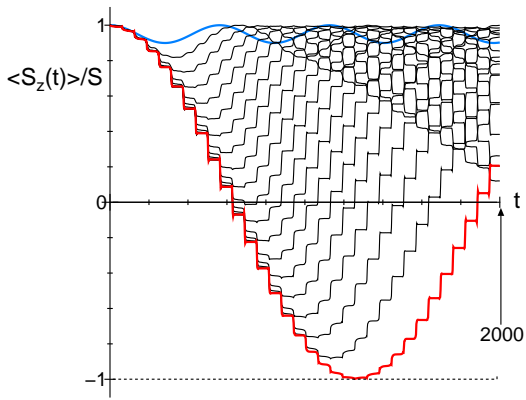}
\end{array}
$$
\caption{$\langle S_z(t) \rangle$ for various values of $S'(=S,S-1,\cdots -S+1)$ for a spin ($S=10$) under the static field $H_z=0.1$ with the amplitude $h_{\rm ac}=0.005$.
$\langle S_z(t) \rangle$(bold curve) normalized by $S$.
}
\label{fig:hmPMZ}
\end{figure}

Interestingly, as we show in the main text, if we add the ac-field up to $S'$ (Eq.~(6) in the main text) ,
$$
{\cal H}=-DS_z^2-H_zS_z
$$ \beq
-h\left(\sum_{m=S}^{S'}\sin(\omega_{m\rightarrow m-1} t)S_x
+\sum_{m=S}^{S'}\cos(\omega_{m\rightarrow m-1} t)S_y\right).
\eeq
the initial state $|S\rangle$ is transferred to $|S'-1\rangle$.
In Fig.~\ref{fig:hmPMZ}, $\langle S_z(t)\rangle$ shows 
approximate oscillations between $S$ and $S'-1$.
It should be noted that $\langle S_z(t)\rangle=S$ does not spontaneously relax to $-S$ 
even if $S'-1<0$, i.e., it is on the other side of the barrier.

\section{A protocol for the spin reversal}

The case of $S'=-S+1$, including all the resonance fields, the Hamiltonian becomes
\beq
{\cal H}=-DS_z^2-H_zS_z-h_{\rm ac}
\left(\left(\sum_{m=S}^{-S+1}\sin(\omega_{m\rightarrow m-1} t)\right)S_x+
\left(\sum_{m=S}^{-S+1}\cos(\omega_{m\rightarrow m-1} t)\right)S_y\right).
\label{full-ac}
\eeq
In this case, we find almost full reversal from $S$ to $-S$ as we depicted in Fig.~\ref{fig:hmPMZ}.
This is similar to the usual Rabi oscillation (Fig.~2), but the spin-fidelity 
is not constant (see Fig.7 in the main text), simply because we have quantum and not classical Rabi oscillations. Thus, physical natures of the oscillation are different.

\subsection{Compact form of the sum of ac-fields and $H_z$ dependence}\label{secIIIA}
\begin{figure}[h]
$$
\begin{array}{c}
\includegraphics[width=6cm]{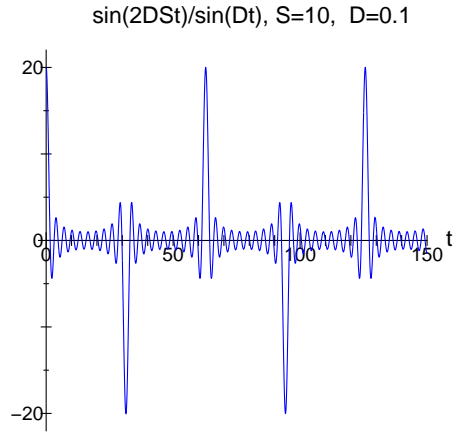} 
\end{array} 
$$
\caption{The shape of $f(t)$ for $S=10$ and $D=0.1$.}
\label{f-dependence}
\end{figure}

The sum of the time-dependent part is written as
$$
\sum_{m=S}^{-S+1}\cos(\omega_{m\rightarrow m-1} t)=\sum_{m=S}^{-S+1}{e^{i\omega_{m\rightarrow m-1} t}+e^{-i\omega_{m\rightarrow m-1} t}\over 2}
=\sum_{m=S}^{-S+1}{e^{i(H_z+D(2m-1)) t}+e^{-i(H_z+D(2m-1) t}\over 2}
$$ $$
={1\over 2}\left[
e^{i(H_z-D)t}\sum_{m=S}^{-S+1}e^{2iDmt}+e^{-i(H_z-D)t}\sum_{m=S}^{-S+1}e^{-2iDmt}
\right]
$$ $$
={1\over 2}\left[
e^{i(H_z-D)t}{e^{2iSDt}-e^{-2iSDt}\over1-e^{-2iDt}}+e^{-i(H_z-D)t}{e^{-2iSDt}-e^{2iSDt}\over1-e^{2iDt}}
\right]
$$ \beq
={1\over 2}\left[
e^{iH_z t}{e^{2iSDt}-e^{-2iSDt}\over e^{iDt}-e^{-iDt}}+e^{-iH_zt}{e^{-2iSDt}-e^{2iSDt}\over e^{-iDt}-e^{iDt}}.
\right]
\eeq
Introducing the function:
\beq
f(t)\equiv{e^{2iSDt}-e^{-2iSDt}\over e^{iDt}-e^{-iDt}}={\sin(2DSt)\over\sin(Dt)},
\label{ft}
\eeq
the sum becomes
\beq
\sum_{m=S}^{-S+1}\cos(\omega_{m\rightarrow m-1} t)=
{1\over 2}\left[f(t)\left(e^{iH_z t}+e^{-iH_zt}\right)\right]=f(t)\cos(H_z t).
\eeq
In the same way, 
\beq
\sum_{m=S}^{-S+1}\sin(\omega_{m\rightarrow m-1} t)=\sum_{m=S}^{-S+1}{e^{i\omega_{m\rightarrow m-1} t}-e^{-i\omega_{m\rightarrow m-1} t}\over 2i}
=
{1\over 2i}\left[f(t)\left(e^{iH_z t}-e^{-iH_zt}\right)\right]=f(t)\sin(H_z t).
\eeq
We plot the shape of $f(t)$ in Fig.~\ref{f-dependence} for $S=10$ and $D=0.1$.

Using the relations, 
\beq
\left(\sum_{m=S}^{-S+1}\sin(\omega_{m\rightarrow m-1} t)\right)S_x+
\left(\sum_{m=S}^{-S+1}\cos(\omega_{m\rightarrow m-1} t)\right)S_y
={f(t)\over 2i}\left(S^+e^{iH_z t}-S^-e^{-iH_z t}\right).
\eeq
The \SDG is given by
\beq
i\hbar{\partial \over\partial t}|\Psi(t)\rangle={\cal H}|\Psi(t)\rangle
=\left(-DS_z^2-H_zS_z-h{f(t)\over 2i}\left(S^+e^{iH_z t}-S^-e^{-iH_z t}\right)
\right)|\Psi(t)\rangle.
\label{SDGeqftpsi}
\eeq
In the rotating frame:
\beq
|\Psi(t)\rangle=e^{iH_zS_z{t}/\hbar}|\Phi(t)\rangle,
\eeq
The \SDG becomaes
$$
i\hbar\left(iH_zS_z/\hbar e^{iH_zS_z{t}/\hbar}|\Phi(t)\rangle+e^{iH_zS_z{t}/\hbar}{\partial \over\partial t}
|\Phi(t)\rangle\right)=-H_zS_z+i\hbar e^{iH_zS_z{t}/\hbar}{\partial \over\partial t}
|\Phi(t)\rangle
$$ \beq
=\left(-DS_z^2-H_zS_z-h{f(t)\over 2i}\left(S^+e^{iH_z t}-S^-e^{-iH_z t}\right)
\right)e^{iH_zS_z/\hbar}|\Phi(t)\rangle,
\eeq
and with the relations:
\beq
e^{-iH_zS_z{t}/\hbar}S^+e^{iH_zS_z{t}/\hbar}=e^{-iH_z t}S^+,\quad
e^{-iH_zS_z{t}/\hbar}S^-e^{iH_zS_z{t}/\hbar}=e^{iH_z t}S^-,
\eeq
we finally obtain the \SDG in the rotating frame:
\beq
i\hbar{\partial \over\partial t}|\Phi(t)\rangle=
\left(-DS_z^2-h_{\rm ac}{f(t)\over 2i}\left(S^+-S^-\right)\right)|\Phi(t)\rangle
=\left(-DS_z^2-h_{\rm ac}f(t)S_y\right)|\Phi(t)\rangle.
\label{dphidtft}
\eeq

\begin{figure}[h]
$$
\begin{array}{ccc}
\includegraphics[width=4.5cm]{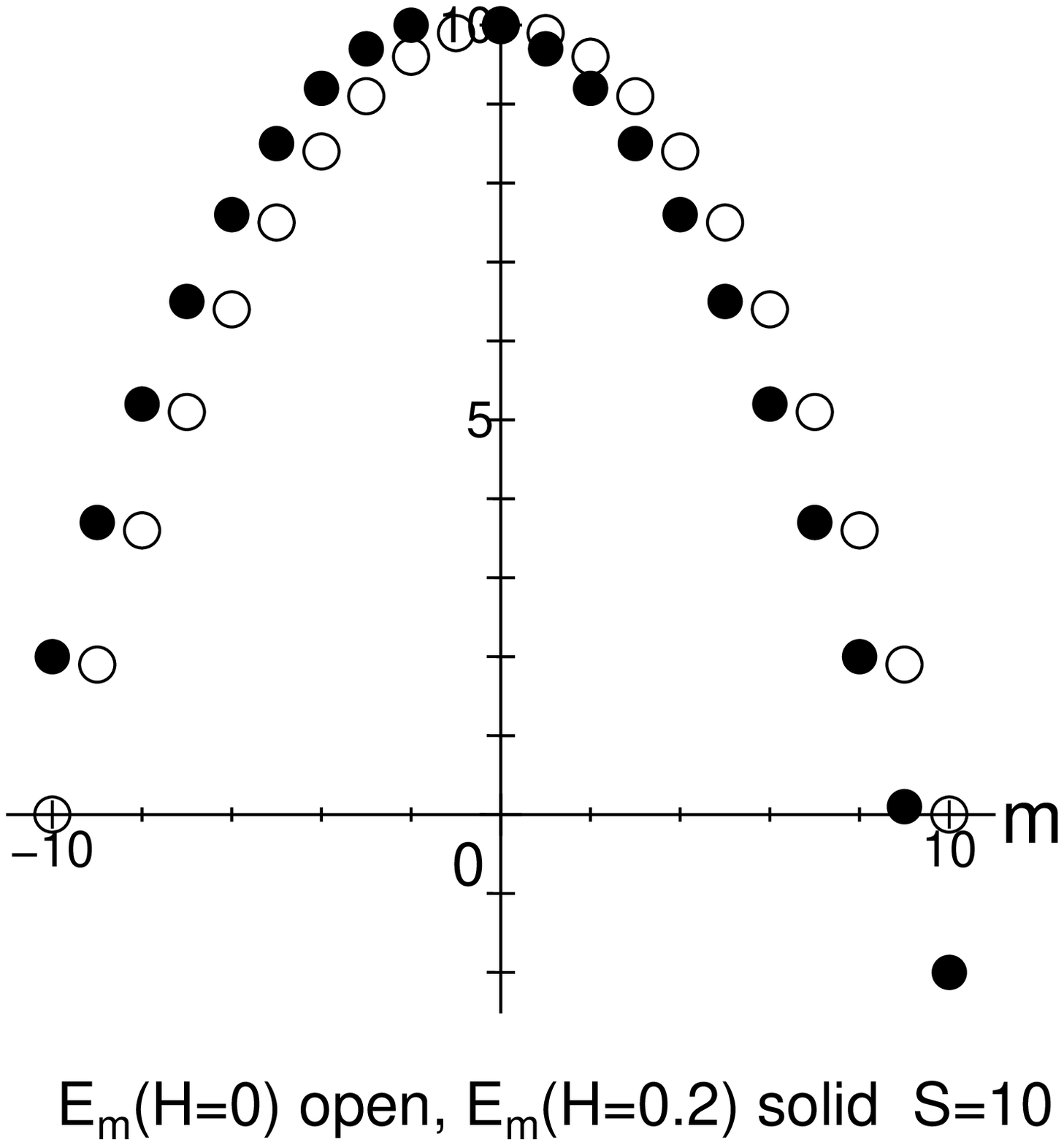} &\quad&
\includegraphics[width=5.5cm]{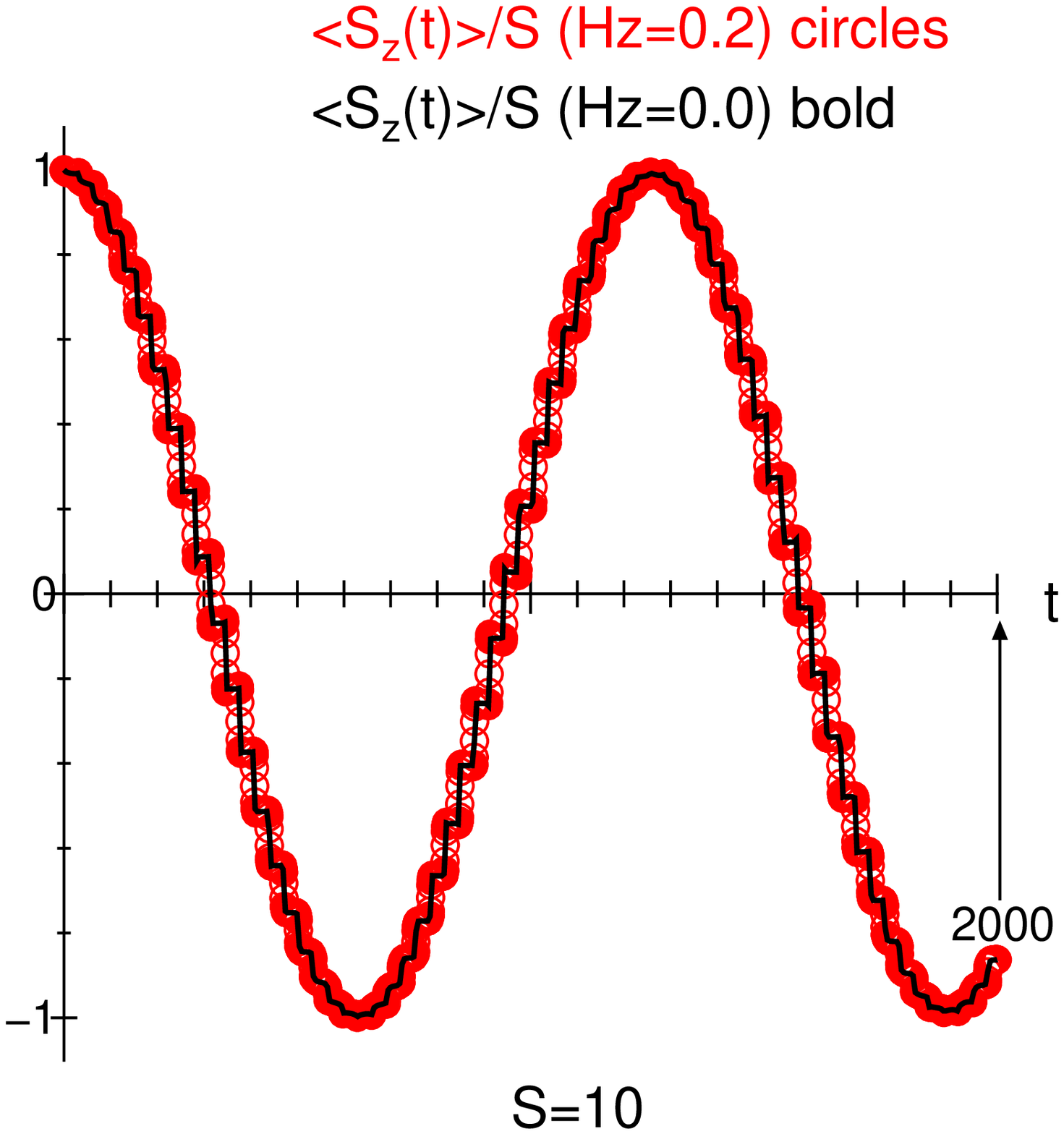}\\
({\rm a}) && ({\rm b})
\end{array} 
$$
\caption{(a) Plot of the reduced energy $E_m-DS^2$ vs $S_z=m$ for $H=0$ (open circle) and $H=0.2$ (closed circle).
(b)
$H_z$ dependence of $S_z(t)/S$ (black bold line) for $S=10$ and the initial state is 
$(S_x,S_y,S_z)=(0,0,1)$.
$H_z=0$ (black curve) and $H_z=0.2$(red circles).
They completely coincide.}
\label{Hz-dependence}
\end{figure}

It should be noted that this \SDG does not depend on $H_z$.
Thus, the dynamics of $\langle S_z(t)\rangle$ is independent of $H_z$ and levels resonance is not required. 
In Fig.~\ref{Hz-dependence}, we compare the cases of $H_z=0$ (black curve) and $H_z=0.2$(red circles).
In the case of $H_z=0$, $S_z=\pm 10$ are degenerate, while in case of $H_z=0.2$, $S_z=10$ is the true stable state and $S_z=-10$ is metastable. 
But $\langle S_z(t)\rangle$ coincides with each other as 
this should be according to (\ref{dphidtft}).

\subsection{$S$-dependence of the period of magnetization reversal}

\begin{figure}[h]
\begin{center}
\includegraphics[width=5cm]{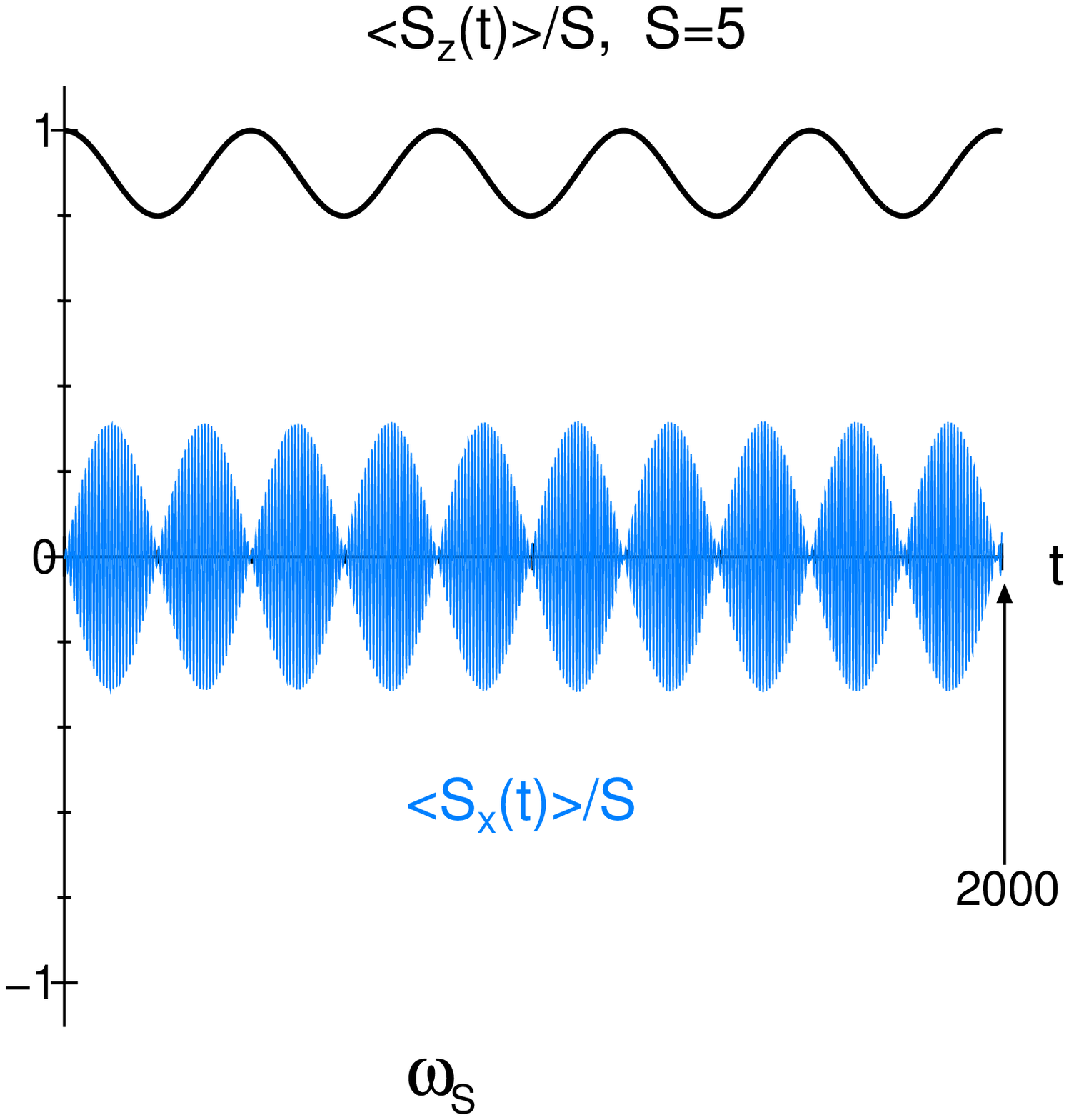}
\includegraphics[width=5cm]{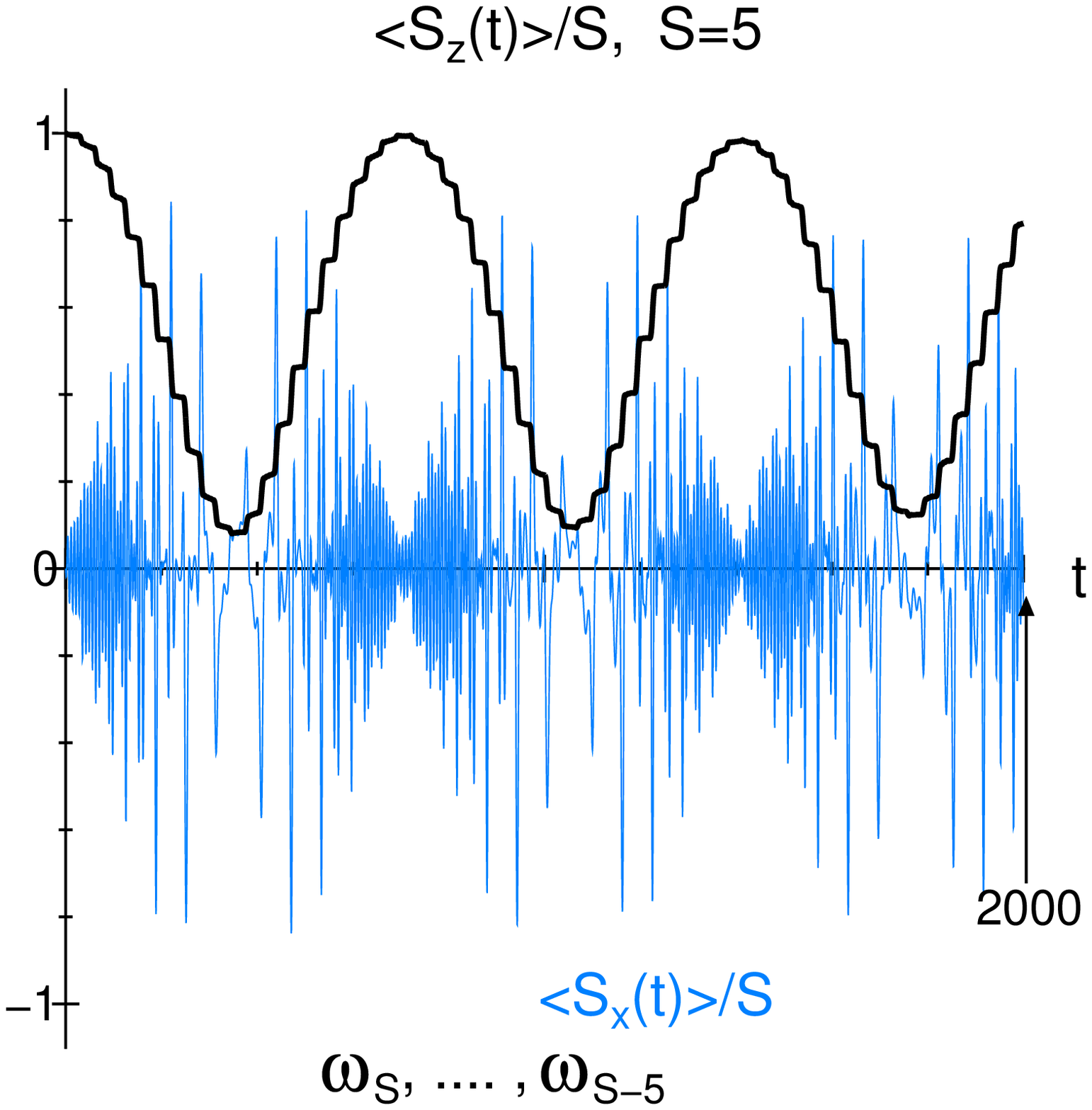}
\includegraphics[width=5cm]{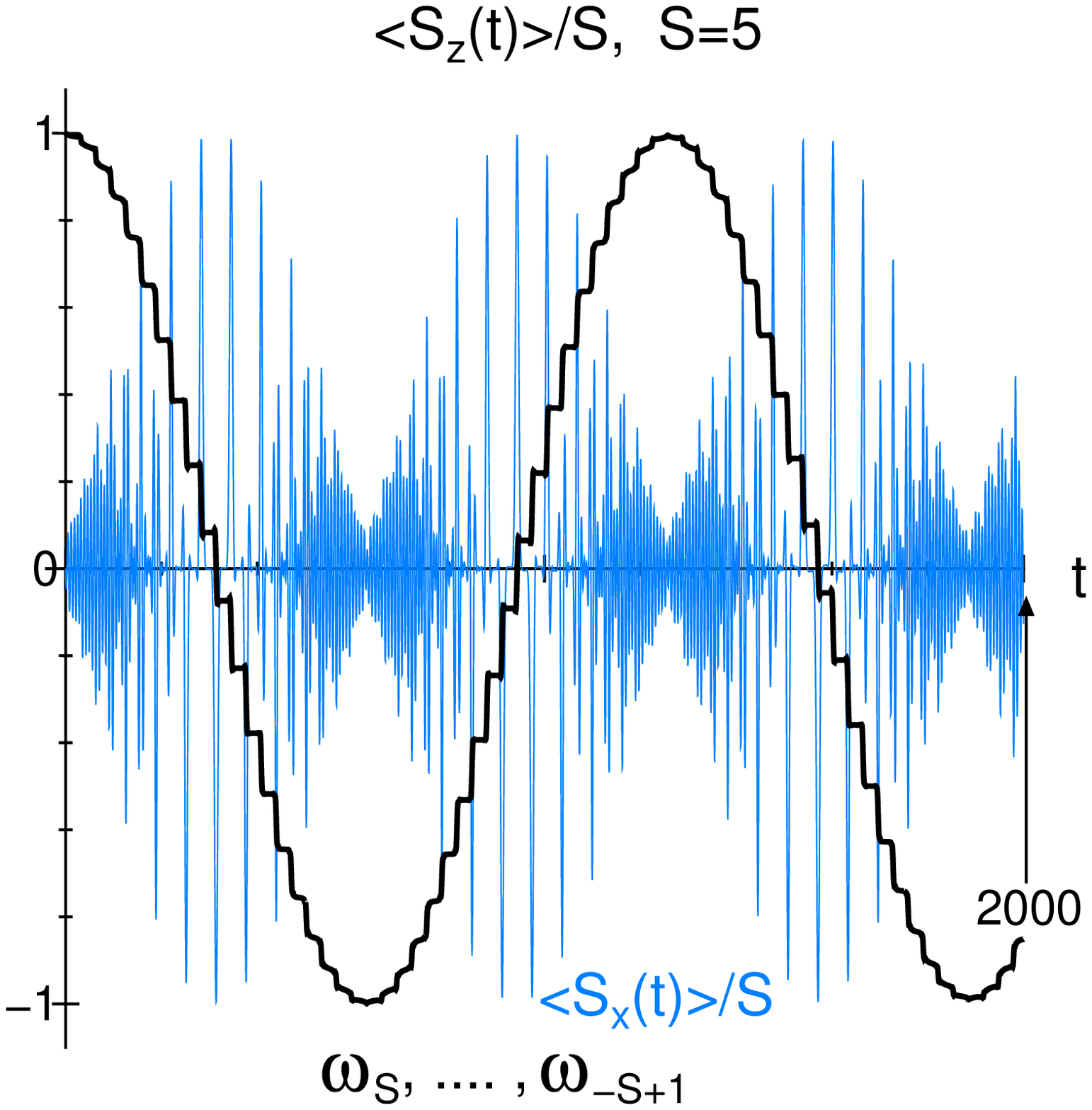}
\end{center}
\caption{Time dependence of magnetizations $S_x(t)/S$ (blue thin line) and $S_z(t)/S$ (black bold line) for $S=5$ under different sequences of ac-fields with $S'=S, S-5$ and $-S+1$. $h=0.005$. The initial state is $(S_x,S_y,S_z)=(0,0,1)$.}
\label{Rabi-S5}
\end{figure}

In this section, we study $S$-dependence of GQOAB.
Starting with the case of different integer values, we show $\langle S_z\rangle$ for $S=5$ in Fig.~\ref{Rabi-S5}.
It is rather surprising that the shape of magnetization reversal is almost the same for $S=10$ and $S=5$.

\begin{figure}[h]
\begin{center}
\includegraphics[width=5cm]{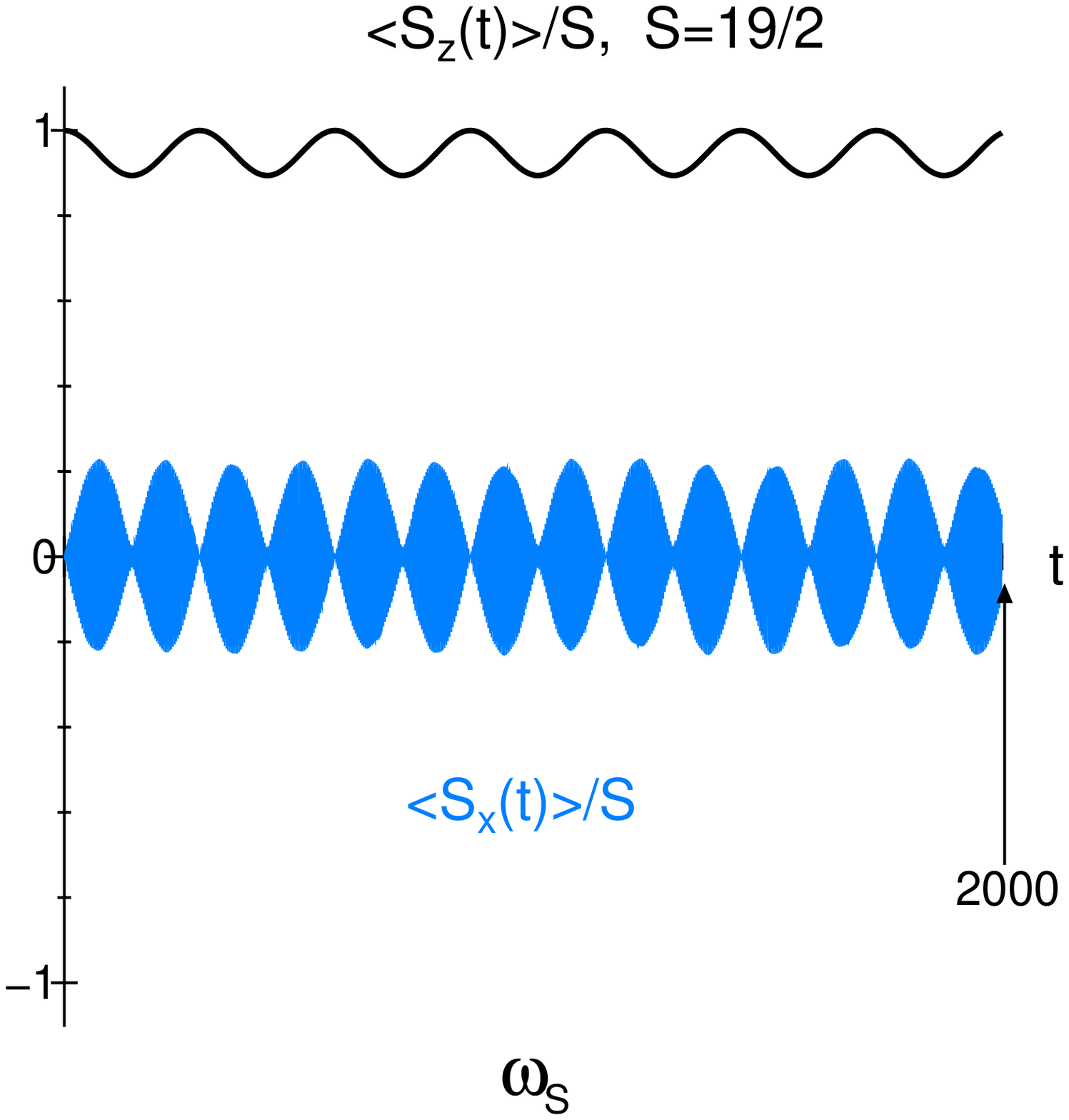}
\includegraphics[width=5cm]{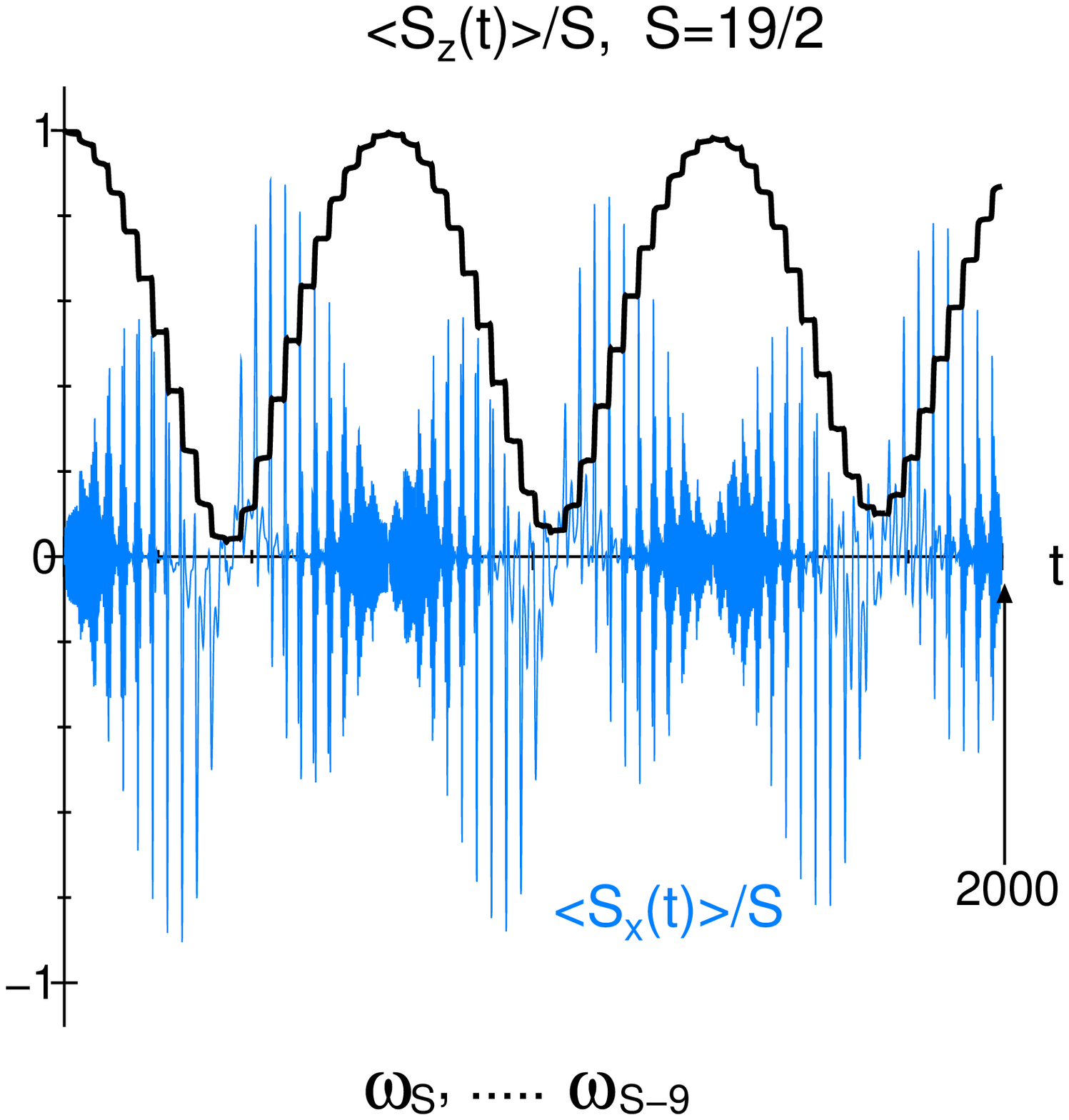}
\includegraphics[width=5cm]{RabiS19multiRFm19.eps}
\end{center}
\caption{Time dependence of magnetizations $S_x(t)/S$ (blue thin line) and $S_z(t)/S$ (black bold line) for $S=19/2$ under different sequences of ac-fields with $S'=S, S-9$ and $-S+1$. $h=0.005$. The initial state is $(S_x,S_y,S_z)=(0,0,1)$.}
\label{Rabi-S19}
\end{figure}

Next, we study the case of half-integer spin $S=19/2$.
In the case of integer $S$, the maximum energy is at $S_z=0$, while in the case of half-integer $S$, there are two states of maximum energy, i.e., $S_z=\pm 1/2$. Thus, we may expect some difference.
However, it is found that the shape of magnetization reversal is almost the same as those of $S=10$ and $S=19/2$ as shown in Fig.~\ref{Rabi-S19}.

\subsection{Dependence of the period of magnetization reversal on the amplitude of $f(t)$}
\label{sec:ft-dependence}

\begin{figure}[h]
\begin{center}
\includegraphics[width=5cm]{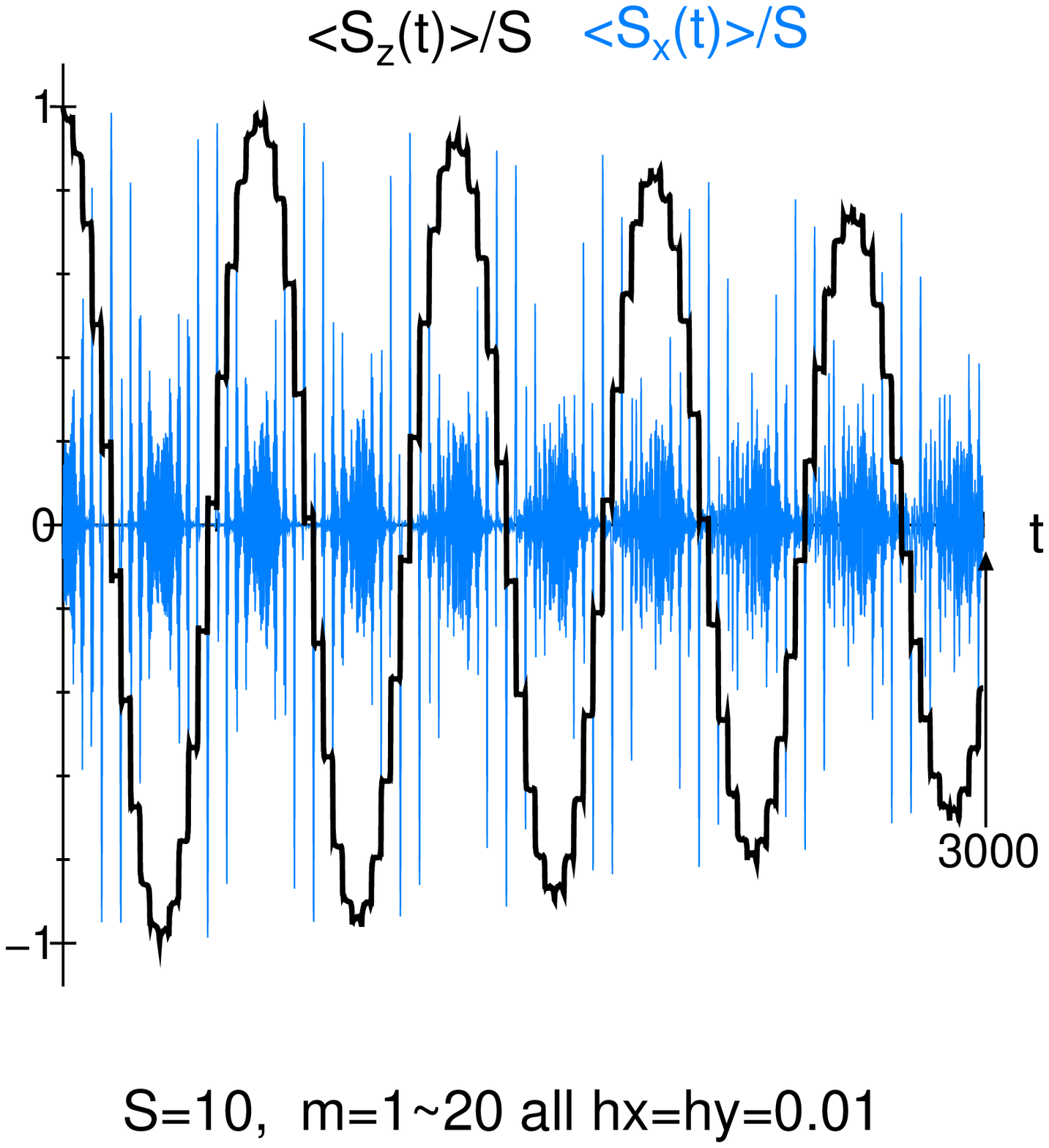}
\includegraphics[width=5cm]{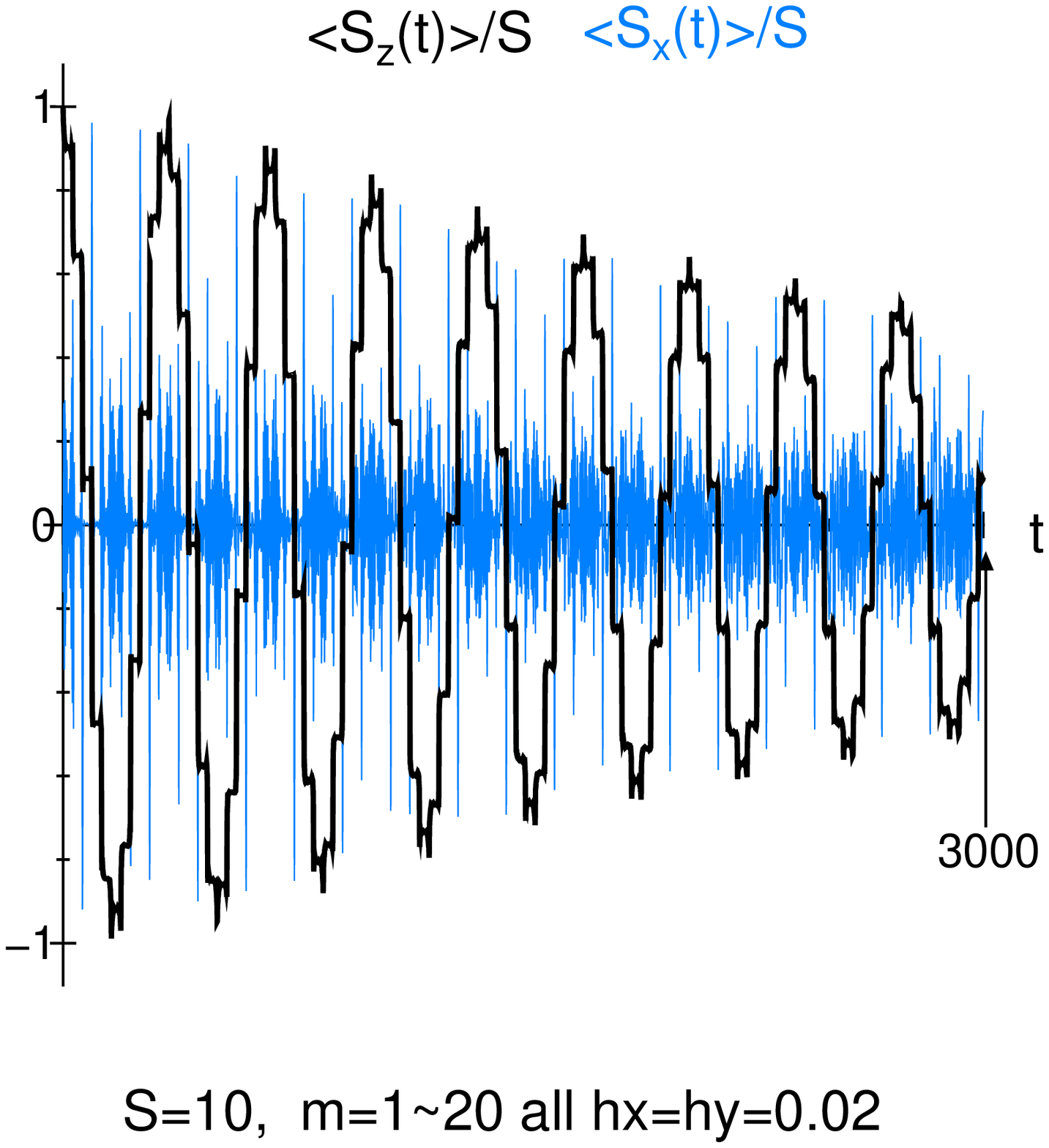}
\end{center}
\caption{Time dependence of $S_x(t)/S$ (blue thin line) and $S_z(t)/S$ (black bold line) for $S=10$ under the AC field of $\omega_{S\rightarrow S-1}$ only. $h_{\rm ac}=0.01$ and 0.02. 
The initial state is $(S_x,S_y,S_z)=(0,0,1)$.}
\label{h-depS10}
\end{figure}

So far, we studied the effect of $f(t)$, the amplitude of which being given by $h_{\rm ac}=0.005$.
In this section, we study the period of magnetization reversal on $h_{\rm ac}$.
In Fig.~\ref{h-depS10}, we depict the cases with $h_{\rm ac}=0.01$ and 0.02.
Here we find that the period is almost inversely proportional to $h_{\rm ac}$.
This dependence is the same as 
the one of the original Rabi oscillation $T=2\pi/h_{\rm ac}$.
The energy structure and the driving filed $f(t)$ depend on $S$, but the time evolutions of $\langle S_z(t)\rangle$ seem the same. Thus, $f(t)$ plays the role of the resonance ac-field for the original Rabi oscillation for the free spin in GQOAB. 

\subsection{$D$ dependence}\label{sec:D-dependence}

$D$ dependence is rather tricky. When $D$ increases, the resonance frequency $\omega_{m\rightarrow m-1}=(2m-1)D$ increases. Thus, the frequencies for ac-field increase.
However, the period of GQOAB is given only by $h_{\rm ac}$ and does not depend on $D$.
To see this fact, let is set $Dt=\tau$ in the equation (\ref{dphidtft}). Then, we have
\beq
iD\hbar{\partial \over\partial \tau}|\Phi(t)\rangle=
\left(-DS_z^2-h_{\rm ac}{f(t)\over 2i}\left(S^+-S^-\right)\right)|\Phi(t)\rangle
=\left(-DS_z^2-h_{\rm ac}f(t)S_y\right)|\Phi(t)\rangle.
\eeq
and thus
\beq
i\hbar{\partial \over\partial \tau}|\Phi(t)\rangle
=\left(-S_z^2-{h_{\rm ac}\over D}f(\tau)S_y\right)|\Phi(t)\rangle, \quad f(\tau)={\sin(S\tau)\over\sin\tau}.
\label{dphidtftau}
\eeq
In the time $\tau$, the GQOAB occurs with the amplitude $h_{\rm ac}/D$
and a period of oscillations proportional to
$1/({h_{\rm ac}/D})$, that is, the period increases proportionally to $D$ in the time $\tau$ (see Sec.~\ref{sec:ft-dependence}).
Therefore, in the original time scale $t$ the period does not depend on $D$.

\begin{figure}[h]
$$
\begin{array}{ccc}
\includegraphics[width=5cm]{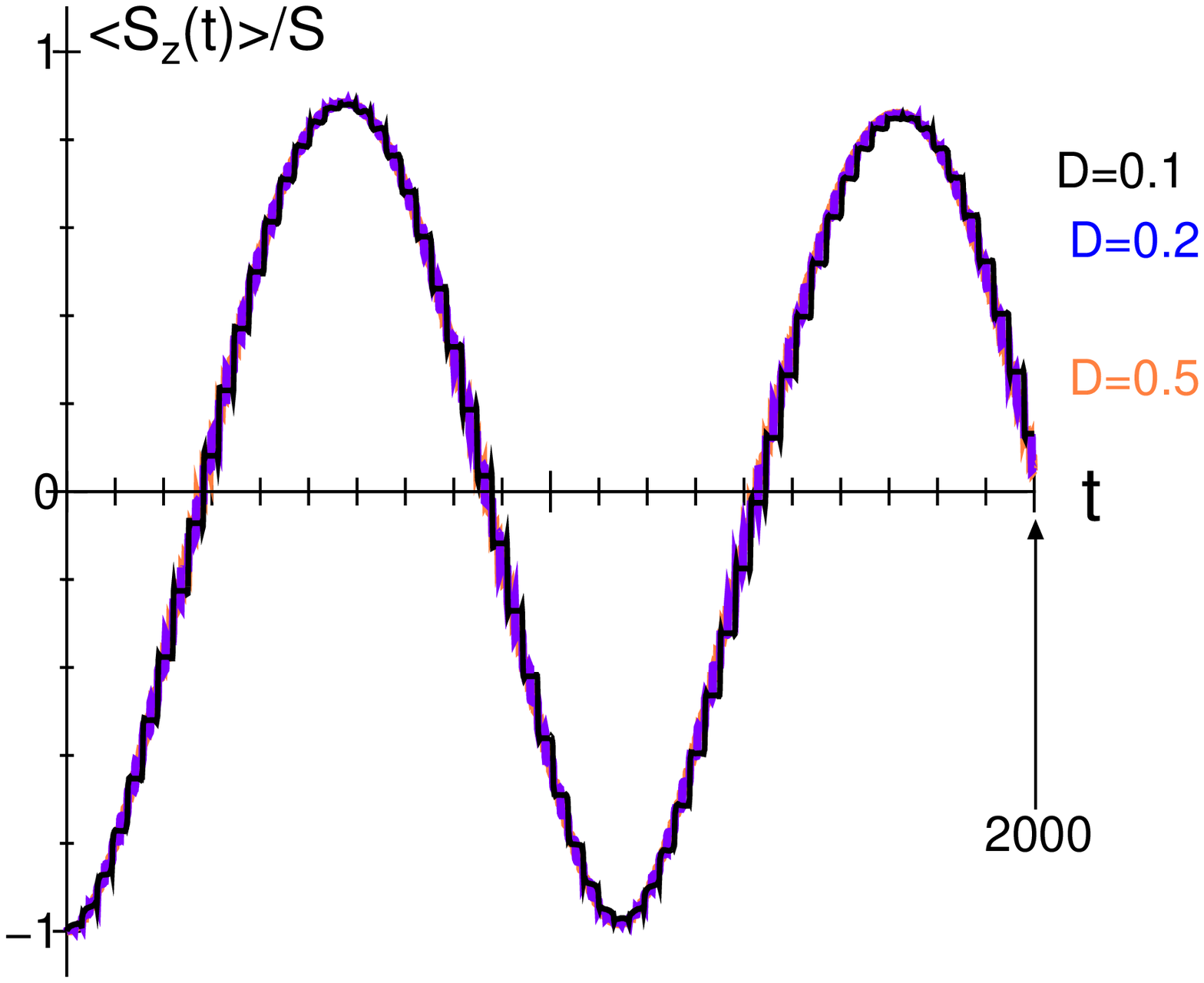}&
\includegraphics[width=5cm]{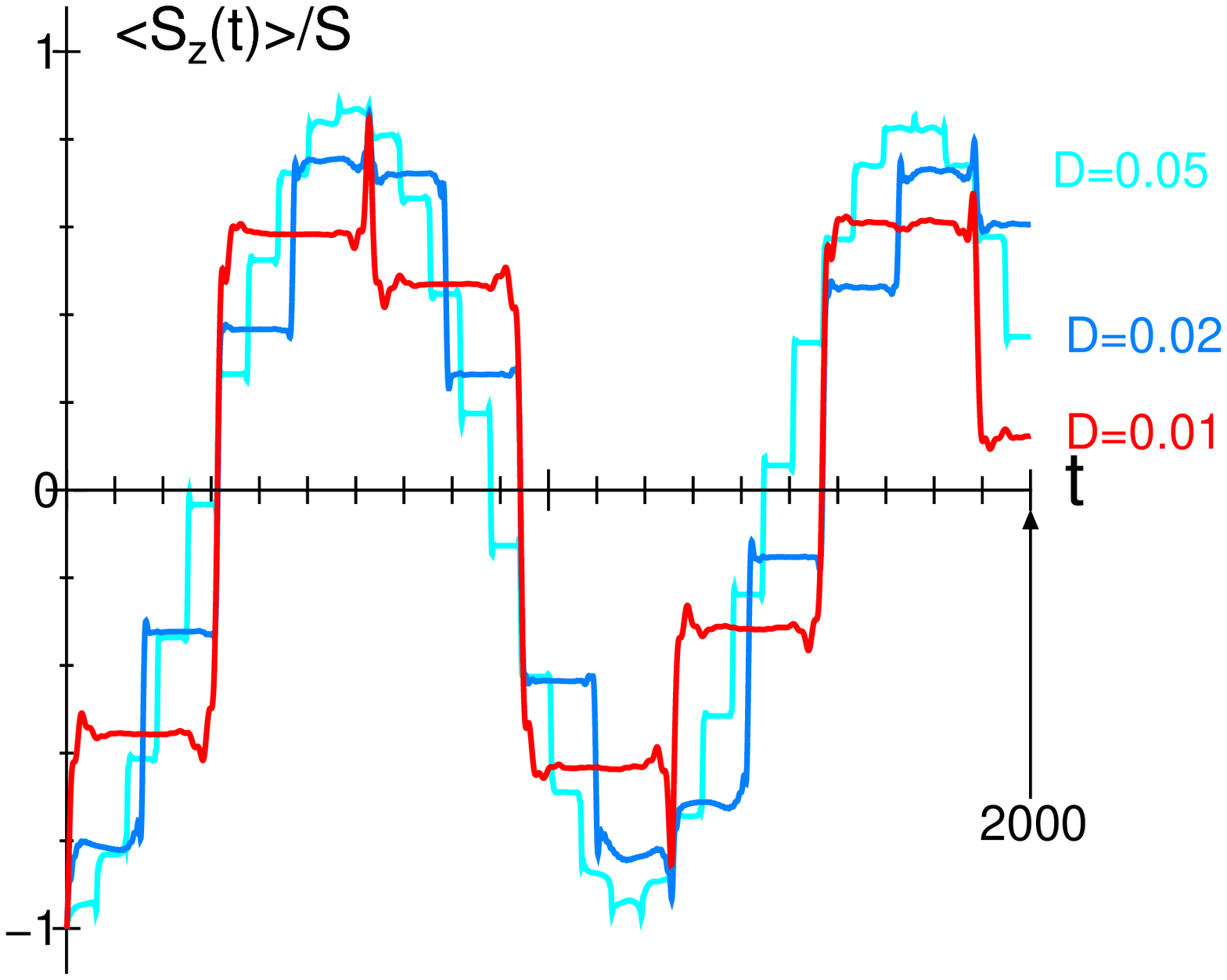}&
\includegraphics[width=5cm]{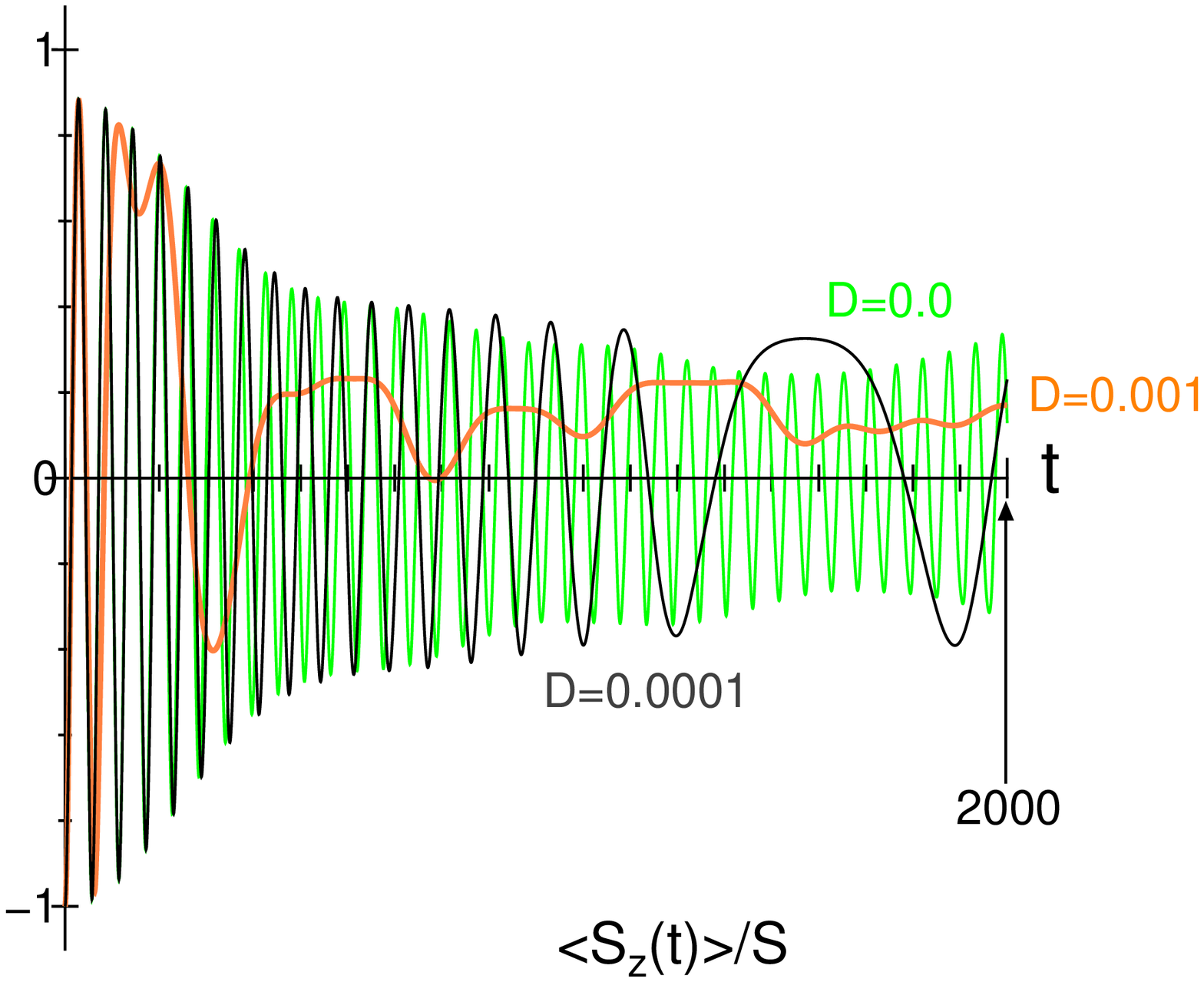}\\
({\rm a})&({\rm b})&({\rm c})
\end{array}
$$
\caption{$D$-dependence of the time dependence of $S_x(t)/S$ (blue thin line) and $S_z(t)/S$ (black bold line) for $S=10$ with all the AC field and $H_z=0.0$. The initial state is $(S_x,S_y,S_z)=(0,0,1)$.
(a) $D=0.5$, 0.2, and 0.1, (b) $D=0.05$, 0.02 and 0.01, (c) $D=0.001$, 0.0001 and 0.}
\label{fig:D-depS10}
\end{figure}

However, we noted that the dynamics becomes more complex when $D$ approaches to 0
as shown in Fig.~\ref{fig:D-depS10}.
More particularly,
if $D=0$ and $H_z\ne 0$, the initial Hamiltonian goes back to the one of usual of Rabi oscillations. 
And so, when $D\rightarrow 0$, the period of oscillation changes from
$2\pi/h_{\rm ac}$ (GQOAB case) to $2\pi/(2Sh_{\rm ac})$ (Rabi case). 
In between the dynamics becomes complex, as shown in Fig.~\ref{fig:D-depS10}, due to a continuous transition between both types of oscillations.


\begin{thebibliography}{99}%
\bibitem{SMM} L. Thomas, F. Lionti, R. Ballou, D. Gatteschi, R. Sessoli and B. Barbara: Nature {\bf 383}, 145 (1996).
J. R. Friedman, M. P. Sarachik, J. Tejada, and R. Ziolo, Phys. Rev. Lett. {\bf 76}, 3830 (1996).
J. A. A. J. Perenboom, J. S. Brooks, S. Hill, T. Hathaway, and N. S. Dalal., Phys. Rev. B {\bf 58}, 330 (1998).
\bibitem{SMM-Fe8} W. Wernsdorfer, R. Sessoli, A. Caneschi, D. Gatteschi, A. Cornia, and D. Mailly, J. Appl. Phys. {\bf 87}, 5481 (2000),
E. del Barco, J. M. Hernandez, J. Tejada, N. Biskup, R. Achey, I. Rutel, N. Dalal, and J. Brooks,
Phys. Rev. B {\bf 62}, 3018, (2000).
%
\bibitem{Lanthanide1} N. Ishikawa, M. Sugita, T. Ishikawa, S. Koshihara, and Y. Kaizu,
J. Am. Chem. Soc. {\bf 125}, 8694 (2003).
\bibitem{Lanthanide2} N. Ishikawa, M. Sugita, and W. Wernsdorfer,
J. Am. Chem. Soc. {\bf 127}, 3650 (2005).
%
\bibitem{SMM2} D. Gatteschi, R. Sessoli, and J. Villain, {\it Molecular Nanomagnets}, Oxford Univ. Press 2006.

\bibitem{Sorace} L. Sorace, W. Wernsdorfer, C. Thirion, A.-L. Barra, M. Pacchioni, D. Mailly, and B. Barbara,
Phys. Rev. B {\bf 68}, 220407(R) (2003).
%

\bibitem{sec:SM} Supplementary material. [Attached below]

\bibitem{Hatomura2016}T. Hatomura, B. Barbara and S. Miyashita,
Phys. Rev. Lett. {\bf 116} (2016) 037203.
\bibitem{SM2022} S. Miyashita, {\it Collapse of Metastability}, Springer-Nature (2022).

\bibitem{YLiF4} R. Giraud, W Wernsdorfer, A M Tkachuk, D Mailly, and B Barbara,
Phys. Rev. Lett, {\bf 87}, 057203 (2001).

\bibitem{NN}%Perspectives in Magnetic Resonance Perspectives of shaped pulses for EPR spectroscopy
P. E. Spindler, Philipp Sch\"{o}ps, W. Kallies, S. J. Glaser, T. F. Prisner,
J. Mag. Res. {\bf 280}, 30 (2017).

\bibitem{Y} B. Barbara, Mesoscopic systems; classical irreversibility and quantum coherence,
Phil. Trans. R. Soc., {\bf 370}, 4487 (2012).

\bibitem{N6} C. Bonizzoni, A. Ghirri, K. Bader, J. van Slageren, M. Perfetti, L. Sorace,
Y. Lan, O. Fuhr, M. Rubene, and M. Affrontea, Dalton Trans., {\bf 45}, 16596 (2016).
%Coupling molecular spin centers to microwave planar resonators: towards integration of molecular qubits in quantum circuits
\bibitem{N7} C. Bonizzoni, A. Ghirri, F. Santanni, M. Atzori, L. Sorace, R. Sessoli and
M. Affronte, npj Quantum Information 68, (1-7) (2020). %Nature Commun. s41534-020
%Storage and retrieval of microwave pulses with molecular spin ensembles
\bibitem{N8} I. Gimeno, A. Urtizberea, J. Roman-Roche, 
D. Zueco, A. Camon, P. J. Alonso, O. Roubeau and F. Luis,
Chem. Sci. {\bf 12}, 5621 (2021).
%Broad-band spectroscopy of a vanadyl porphyrin: a model electronuclear spin qudit
\bibitem{N9} M. D. Jenkins, Y. Duan, B. Diosdado, J. J. Garcia-Ripoll, A. Gaita-Arino, C. Gimenez-Saiz, P. J. Alonso, E. Coronado, and F. Luis, Phys. Rev. B {\bf 95}, 064423 (2017).
%Coherent manipulation of three-qubit states in a molecular single-ion magnet Gd3+, Y3+ S=7/2
\bibitem{N10} V. Rollano, M. C. de Ory, C. D. Buch, M. Rubin-Osanz, D. Zueco,
C. Sanchez-Azqueta, A. Chiesa, D. Granados, S. Carretta,
A. Gomez, S. Piligkos and F. Luis, Commun. Phys., {\bf 5:246} (1-9) (2022). % s42005
%High cooperativity coupling to nuclear spins on a circuit quantum electrodynamics architecture
\bibitem{N11} G. Franco-Rivera, J. Cochran, L. Chen, S. Bertaina, and I. Chiorescu,
Phys. Rev. Applied {\bf 18}, 014054.
%On-Chip Detection of Electronuclear Transitions in the 155,157Gd Multilevel Spin

\bibitem{N12} M. Chizzini, L. Crippa, L. Zaccardi, ad E. Macaluso, 
S. Carretta, A. Chiesa, and P. Santini, Phys. Chem. Chem. Phys.,
{\bf 24}, 20030 (2022).
%Quantum error correction with molecular spin qubits
\bibitem{N13} S. Carretta, D. Zueco, A. Chiesa, A. Gomez-Leon, and F. Luis,
Appl. Phys. Lett. {\bf 118}, 240501 (2021).
%A perspective on scaling up quantum computation with molecular spins
\bibitem{N14} S. Chicco, A. Chiesa, G. Allodi, E. Garlatti,
M. Atzori, L. Sorace, R. De Renzi, R. Sessoli, and S. Carretta,
Chem. Sci., {\bf 12}, 12046 (2021).
%Controlled coherent dynamics of [VO(TPP)], a prototype molecular nuclear qudit with anelectronic ancilla
\bibitem{N15} E. Macaluso, M. Rubin, D. Aguilia, A. Chiesa, 
Leoni A. Barrios, ef J. I. Martinez, P. J. Alonso, cd O. Roubeau,
F. Luis, G. Aromi, and S. Carretta, Chem. Sci., {\bf 11}, 10337 (2020).
%A heterometallic [LnLn0Ln] lanthanide complex as a qubit with embedded quantum error correction
\bibitem{N16} A. Castro, A. Garcia Carrizo, S. Roca, D. Zueco, and
F. Luis, Phys. Rev. Applied {\bf 18}, 064028 (2022).
%Optimal Control of Molecular Spin Qubits

\bibitem{N2} C. Godfrin, A. Ferhat, R. Ballou, S. Klyatskaya, M. Ruben, W. Wernsdorfer and F. Balestro, Phys. Rev. Lett. {\bf 119}, 187702 (2017). 

\bibitem{CL} A. J. Leggett, S. Chakravarty, A. T. Dorsey, M. P. A. Fisher, A. Garg, and W. Zwerger, Rev. Mod. Phys. {\bf 59},1 (1987).
A. O. Caldeira and A. J. Leggett, Phys. Rev. Lett. {\bf 46}, 211 (1981).
A. O. Caldeira and A. J. Leggett, Ann Phys. {\bf 149}, 374 (1983).
A. O. Caldeira and A. J. Leggett. Phys. Rev. A{\bf 31}, 1059 (1985).
\bibitem{Spinbath} %Theory of the spin bath
N. V. Prokof'ev and P. C. E. Stamp, Rep. Prog. Phys. {\bf 63}, 669 (2000).

\bibitem{Z1} %Quantum oscillations in a molecular magnet 
S. Bertaina, S. Gambarelli, T. Mitra, B. Tsukerblat, A. M\"{u}ller, B. Barbara, Nature {\bf 453}, 203 (2008).

\bibitem{Z2} %Coherent spin dynamics in gadolinium-doped CaWO$_4$ crystal
E. I. Baibekov, M. R. Gafurov, D. G. Zverev, I. N. Kurkin, A. A. Rodionov, B. Z. Malkin, and B. Barbara, Phy. Rev. B{\bf 95}, 064427 (2017).

\bibitem{Irinel-Sylvain} H. De Raedt, S. Miyashita, K. Michielsen1, H. Vezin, S. Bertaina, and I. Chiorescu, Eur. Phys. J. B, 95:158 (1-14) (2022).



\end{thebibliography}
\end{document}